\documentclass[11pt,a4paper]{article}
\pdfoutput=1

\usepackage{jheppub}
\usepackage{color}
\usepackage[dvipsnames]{xcolor}
\usepackage{graphicx}
\usepackage{wrapfig}
\usepackage{verbatim}
\usepackage{amsmath}
\usepackage{amssymb}
\usepackage{url}
\usepackage{bbold}
\usepackage{xspace}
\usepackage{slashed}
\usepackage{multirow}
\usepackage{threeparttable}
\usepackage{paralist}
\usepackage{subcaption}
\usepackage{floatrow}
\usepackage{multirow}
\usepackage{arydshln}
\usepackage[normalem]{ulem}

\usepackage{tikz}
\usetikzlibrary{trees}
\usetikzlibrary{decorations.pathmorphing}
\usetikzlibrary{decorations.markings}
\usetikzlibrary{arrows}

\newcommand{\GeV}{\,\mathrm{GeV}}
\newcommand{\TeV}{\,\mathrm{TeV}}
\newcommand{\pb}{\,\mathrm{pb}}
\newcommand{\nb}{\,\mathrm{nb}}
\newcommand{\fb}{\,\mathrm{fb}}
\newcommand{\ab}{\,\mathrm{ab}}

\newcommand{\beq}{\begin{equation}}
\newcommand{\eeq}{\end{equation}}
\newcommand{\bea}{\begin{eqnarray}}
\newcommand{\eea}{\end{eqnarray}}

\newcommand{\Dlr}{~\!\overset{\leftrightarrow}{\hspace{-0.1cm}D}\!}

\DeclareRobustCommand{\Sec}[1]{Sec.~\ref{#1}}

\DeclareRobustCommand{\Tab}[1]{Table~\ref{#1}}

\DeclareRobustCommand{\Fig}[1]{Fig.~\ref{#1}}

\DeclareRobustCommand{\Eq}[1]{Eq.~(\ref{#1})}

\newcommand{\ctt}{c_{tt}}
\newcommand{\cqq}{c_{qq}}
\newcommand{\cqqe}{c_{qq}^{\text{\tiny (8)}}}
\newcommand{\ctq}{c_{tq}}
\newcommand{\ctqe}{c_{tq}^{\text{\tiny (8)}}}
\newcommand{\cht}{c_{Ht}}
\newcommand{\ctD}{c_{tD}}
\newcommand{\ctDe}{c_{tD}^{\text{\tiny (8)}}}
\newcommand{\cqD}{c_{qD}}
\newcommand{\cqDe}{c_{qD}^{\text{\tiny (8)}}}
\newcommand{\cqDt}{c_{qD}^{\text{\tiny (3)}}}
\newcommand{\cte}{c_{te}}
\newcommand{\ctl}{c_{t\ell}}

\newcommand{\Sp}{\mathrm{S}}
\newcommand{\Tp}{\mathrm{T}}
\newcommand{\Zp}{\mathrm{Z}}
\newcommand{\Wp}{\mathrm{W}}
\newcommand{\Yp}{\mathrm{Y}}
\newcommand{\Hp}{\mathrm{H}}

\newcommand{\sigmar}{\sigma^{\text{\tiny SM,\,R}}}

\newcommand{\lhc}{\text{\tiny LHC}}
\newcommand{\fcc}{\text{\tiny FCC}}
\newcommand{\clic}{\text{\tiny CLIC}}
\newcommand{\ilc}{\text{\tiny ILC}}
\newcommand{\fccee}{\text{\tiny FCC-ee}}
\newcommand{\sm}{\text{\tiny SM}}

\notoc

\begin{document}

{\large
\flushright TUM-HEP-1286-20\\ }
\vspace{-0.425cm}
{\large
\flushright CERN-TH-2020-166\\ }

\title{The Present and Future of Four Top Operators}

\author[b]{Giovanni Banelli,}
\author[a]{Ennio Salvioni,}
\author[b]{Javi Serra,}
\author[b]{Tobias Theil,}
\author[b]{Andreas Weiler}

\affiliation[a]{Theoretical Physics Department, CERN, Esplanade des Particules 1, 1211 Gen\`eve 23, Switzerland}
\affiliation[b]{Physik-Department, Technische Universit\"at M\"unchen, James-Franck-Strasse~1, 85748 Garching, Germany}


\date{\today}

\abstract{We study the phenomenology of a strongly-interacting top quark at future hadron and lepton colliders, showing that the characteristic four-top contact operators give rise to the most significant effects. We demonstrate the extraordinary potential of a 100 TeV proton-proton collider to directly test such non-standard interactions in four-top production, a process that we thoroughly analyze in the same-sign dilepton and trilepton channels, and explore in the fully hadronic channel. Furthermore, high-energy electron-positron colliders, such as CLIC or the ILC, are shown to exhibit an indirect yet remarkable sensitivity to four-top operators, since these constitute, via renormalization group evolution, the leading new-physics deformations in top-quark pair production. We investigate the impact of our results on the parameter space of composite Higgs models with a strongly-coupled (right-handed) top quark, finding that four-top probes provide the best sensitivity on the compositeness scale at the future energy frontier. In addition, we investigate mild yet persisting LHC excesses in multilepton plus jets final states, showing that they can be consistently described in the effective field theory of such a new-physics scenario.}

\preprint{}
\maketitle

\newpage
\setcounter{page}{1}
\section{Introduction} \label{sec:intro}

The LHC has lived up to the expectations as a top quark factory. Not only have we learnt about the properties of the top thanks to the variety of processes studied, but also about the extent to which dynamics beyond the SM couples to the top. 

The electroweak hierarchy problem is arguably the strongest motivation for top-philic new physics at the TeV scale. The top quark is singled out in view of its large coupling to the Higgs, which in turn suggests that the new physics has strong couplings to the top sector.
This is particularly the case in composite Higgs models where the Higgs arises as a Nambu-Goldstone boson \cite{Kaplan:1983fs,ArkaniHamed:2002qy,Agashe:2004rs}. The top quark can in fact be regarded as the \emph{raison d'\^etre} of such a scenario, since its large couplings to the composite dynamics are what allows the Higgs field to eventually break the electroweak symmetry. A resulting prime phenomenological implication is that the top exhibits the characteristics of a partially composite state \cite{Kaplan:1991dc}, to the point that in well-motivated realizations of this idea one of its chiral components couples to the new dynamics as strongly as the Higgs does \cite{Giudice:2007fh,Pomarol:2008bh}.
This is the case for instance in composite twin Higgs models and generalizations thereof \cite{Chacko:2005pe,Barbieri:2015lqa,Serra:2017poj}, where consistency with precision data and minimal fine-tuning favor a strongly-coupled $t_R$.

It is of major phenomenological relevance that in these scenarios the anomalous properties of the top sector remain experimentally accessible even if the new resonances are too heavy to be directly produced at colliders. Such anomalous effects are best described by an effective low-energy theory in terms of higher-dimensional operators deforming the SM Lagrangian. In this work we mainly focus on the four-top operator
\beq
\frac{\ctt}{\Lambda^2} (\bar t_R \gamma_\mu t_R) (\bar t_R \gamma^\mu t_R) \, ,
\label{o4tR}
\eeq
since it captures the genuine physics of a strongly-interacting right-handed top quark. Indeed, in the broad class of models we are interested in, this operator is always generated with the largest size and independently of the details of how $t_R$ couples to the new dynamics. 

In addition to the everlasting search for inner structure of the particles that are presently viewed as fundamental, a strong motivation for studying the fate of four-top contact interactions at future colliders is the realization that already at the LHC they give rise to a competitive constraint on the parameter space of composite Higgs (CH) models.
Denoting with $m_*$ the mass of the new heavy degrees of freedom and with $g_*$ their coupling to the Higgs and the (right-handed) top, one expects on dimensional grounds that 
\beq
\frac{\ctt}{\Lambda^2} \sim \frac{g_*^2}{m_*^2} \equiv \frac{1}{f^2} \,.
\label{c4tR}
\eeq
This is of the same size as the operator encoding the inherent deformations associated with a strongly-interacting Higgs field, $O_H = \tfrac{1}{2} (\partial_\mu |H|^2)^2$ with coefficient $c_H/\Lambda^2 \sim 1/f^2$. As we review in \Sec{sec:leplhc}, the current $95\%$ CL constraint on $c_H$ from LEP and LHC Run 2 data translates into $f|_H \gtrsim 820 \GeV$ \cite{Cepeda:2019klc}, only slightly stronger than the bound from $\ctt$: the operator \Eq{o4tR} modifies the rate of four-top production at the LHC with respect to the SM, and the absence of significant deviations yields $f|_{tt} \gtrsim 730 \GeV$ \cite{Aaboud:2018jsj}. Therefore, probes of $O_{tt}$ constitute, already today, a competitive test of CH models.

An added motivation for considering these scenarios is that they generically predict important deviations in processes involving tops and the Higgs or electroweak vector bosons, see e.g.~\cite{Dror:2015nkp}. Intriguingly, present LHC measurements of top-quark pair production in association with a $W$ or $Z$ boson, which are main backgrounds for four-top production, show a consistent pattern of slight excesses with respect to the SM predictions. In \Sec{sec:multilep}, we show their compatibility with precisely the type of heavy new physics we are interested in. Far from a definitive conclusion, we nevertheless find that the excesses are roughly but compellingly consistent with the as yet absence of deviations in other top or Higgs analyses.

We emphasize at this point that while $pp \to t \bar t t \bar t$ is a rare process in the SM, as already noted in \cite{Pomarol:2008bh} the contribution from four-top operators is mainly associated with a strong $t \bar t \to t \bar t$ scattering amplitude that grows with energy, i.e.~$\mathcal{M}_{t \bar t \to t \bar t} \sim \ctt (E/\Lambda)^2$, resulting in an enhanced sensitivity at high-energy colliders.
This high-energy behavior has been exploited in a variety of works exploring new physics in the top sector (e.g.~\cite{Degrande:2010kt,Farina:2012xp,Dror:2015nkp}), and it is behind the spectacular reach of a potential proton-proton collider at a center of mass energy of 100 TeV. We present our analysis of four-top production at the proposed Future Circular Collider (FCC-hh) \cite{Abada:2019lih} in \Sec{sec:hadron}. We concentrate mainly on multilepton final states, although we also explore the fully hadronic signature. We find a sensitivity on the compositeness scale $f|^{\fcc}_{tt} \gtrsim 6.5 \TeV$ at $95\%$ CL after $30 \ab^{-1}$ of integrated luminosity, an order of magnitude above the LHC. The impact on the parameter space $(m_*,g_*)$ of CH models, with a comparison to other relevant probes, is shown in \Fig{fig:FCCbounds}.

At future electron-positron colliders, a direct test of the four-top scattering amplitude  is not feasible. Therefore, the naive expectation would be that the largest sensitivity on top quark compositeness does not arise from \Eq{o4tR}, but from other operators directly modifying $e^+ e^- \to t \bar t$. In \Sec{sec:lepton} we show that this is, in fact, not the case: renormalization group evolution makes the four-top operator dominate the expected size of the effective $(\bar e_{L,R} \gamma_\mu e_{L,R}) (\bar t_R \gamma^\mu t_R)$ operators. The excellent experimental precision achievable at lepton colliders, combined with the possibility of large scattering energies, like the $3$~TeV attainable at the Compact Linear Collider (CLIC)~\cite{Charles:2018vfv}, results in a truly remarkable sensitivity on $\ctt$ and thus on the compositeness scale, e.g.~$f|^{\clic}_{tt} \gtrsim 7.7 \TeV$ for $m_* = 4 \pi f$ at $95\%$ CL and after $3 \ab^{-1}$ of integrated luminosity. As we show in \Fig{fig:LCbounds}, at CLIC these constraints are superior to the corresponding ones from Higgs measurements.

The impact of our results is summarized in Sec.~\ref{sec:conclusions}. For CH models with a strongly-interacting top, both hadronic and leptonic future high-energy colliders will be able to test, through the top quark, fine-tunings of the electroweak scale at the $\xi = v^2/f^2 \approx 10^{-3}$ level, a hundred-fold that of the LHC and certainly unprecedented in the realm of particle physics. This is a truly exceptional motivation for future discoveries that could await us at the high energy frontier.


\section{Operators, expectations and current constraints} \label{sec:leplhc}

\begin{figure}[!t]
    \centering
    \includegraphics[width=0.65\textwidth]{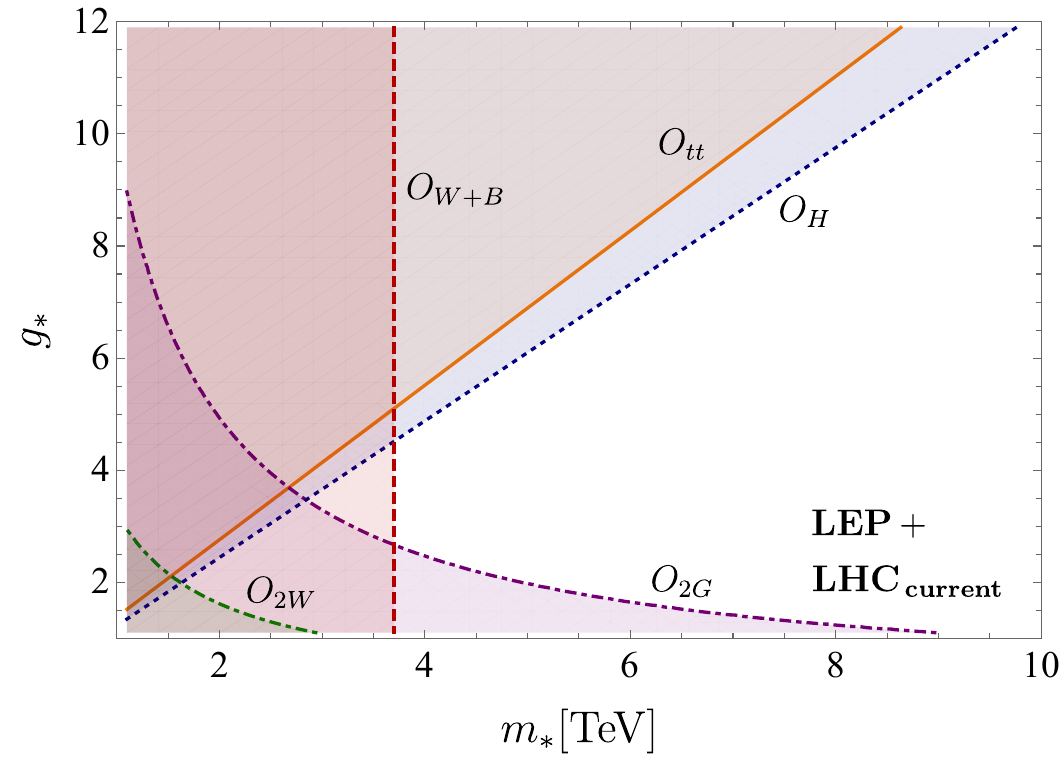}
    \caption{Current 95\% CL excluded regions in the $(m_*,g_*)$ plane of scenarios featuring a strongly-interacting Higgs and (right-handed) top quark. The different limits are associated with bounds on individual operators, each dominating the corresponding observables in a certain region of parameter space (see main text for details) -- this is not a global fit.}
    \label{fig:LHCbounds}
\end{figure}

In this section we define the dimension-six operators we will be considering throughout this work (see \Tab{tab:operators}) and discuss their expected size in theories with a strongly-interacting top quark. We review our current knowledge on the corresponding experimental constraints, paying special attention to those operators leading to the largest sensitivity on the parameter space of CH models. The current status is summarized in \Fig{fig:LHCbounds}.

\renewcommand{\arraystretch}{1.35}
\begin{table}[t]
	\begin{center}
	\begin{tabular}{|c|}
\hline
$O_{tt} = (\bar t_R \gamma_\mu t_R)^2$ \\
\hdashline
$O_{tq} = (\bar t_R \gamma_\mu t_R) (\bar q_L \gamma^\mu q_L)$ \\
$O_{tq}^{\text{\tiny (8)}} = \big(\bar t_R \gamma_\mu t^A t_R \big) \big(\bar q_L \gamma^\mu t^A q_L \big)$ \\
\hdashline
$O_{qq} = (\bar q_L \gamma_\mu q_L)^2$ \\
$O_{qq}^{\text{\tiny (8)}} = \big(\bar q_L \gamma_\mu t^A q_L \big)^2$ \\
\hline
	\end{tabular}
	\hspace{10mm}
	\begin{tabular}{|c|}
\hline
$O_{Ht} = i (H^\dagger \Dlr_\mu H)(\bar t_R \gamma^\mu t_R)$ \\
\hdashline
$O_{Hq} = i (H^\dagger \Dlr_\mu H)(\bar q_L \gamma^\mu q_L)$ \\
$O_{Hq}^{\text{\tiny (3)}} = i (H^\dagger \sigma^a \Dlr_\mu H)(\bar q_L \gamma^\mu \sigma^a q_L)$ \\
\hdashline
$O_{y_t} = y_t H^\dagger H \bar q_L \widetilde{H} t_R$ \\
\hline
	\end{tabular}
	\end{center}
	\begin{center}
	\begin{tabular}{|c|}
\hline
$O_{tD} = (\partial^\mu B_{\mu \nu}) (\bar t_R \gamma^\nu t_R)$ \\
$O_{tD}^{\text{\tiny (8)}} = (D^\mu G_{\mu \nu}^A) (\bar t_R \gamma^\nu t^A t_R)$ \\
\hdashline
$O_{qD} = (\partial^\mu B_{\mu \nu}) (\bar q_L \gamma^\nu q_L)$ \\
$O_{qD}^{\text{\tiny (8)}} = (D^\mu G_{\mu \nu}^A) (\bar q_L \gamma^\nu t^A q_L)$ \\
$O_{qD}^{\text{\tiny (3)}} = (D^\mu W_{\mu \nu}^a) (\bar q_L \gamma^\nu \sigma^a q_L)$ \\
\hline	
	\end{tabular}
	\hspace{10mm}
	\begin{tabular}{|c|}
\hline
$O_H = \tfrac{1}{2} \big( \partial_\mu |H|^2 \big)^2$ \\
$O_T = \tfrac{1}{2} \big( H^\dagger \Dlr_\mu H \big)^2$ \\
\hdashline
$O_W = i g \tfrac{1}{2} (H^\dagger \sigma^a \Dlr_\mu H) D_\nu W^{a \, \mu \nu}$ \\
$O_B = i g' \tfrac{1}{2} (H^\dagger \Dlr_\mu H) \partial_\nu B^{\mu \nu}$ \\
\hdashline
$O_{2G} = -\tfrac{1}{2} (D^\mu G_{\mu \nu}^A)^2$ \\
$O_{2W} = -\tfrac{1}{2} (D^\mu W_{\mu \nu}^a)^2$ \\
$O_{2B} = -\tfrac{1}{2} (\partial^\mu B_{\mu \nu})^2$ \\
\hline
	\end{tabular}
	\end{center}
	\begin{center}
	\begin{tabular}{|c|}
\hline
$\widetilde O_\gamma = H^\dagger H B^{\mu \nu}\widetilde B_{\mu \nu}$ \\
\hline
	\end{tabular}
	\end{center}
	\caption{Set of dimension-six operators relevant in this work, grouped in five different boxes corresponding to the different classes discussed in the main text. Dashed lines within a box separate operators in a given class with a different power counting estimate. We have defined $H^\dagger (\sigma^a) \Dlr_\mu H = H^\dagger (\sigma^a) D_\mu H - (D_\mu H)^\dagger (\sigma^a) H$ and $\widetilde{H} = i \sigma^2 H^\ast$.}
	\label{tab:operators}
\end{table}
\renewcommand{\arraystretch}{1}

Searches for the production of four top quarks at the 13 TeV LHC have provided important constraints on the idea of top quark compositeness. From the absence of significant deviations in the total cross section, which have been searched for using $\approx 36 \fb^{-1}$ of data in the single-lepton \cite{Aaboud:2018xuw,Aaboud:2018jsj,Sirunyan:2019nxl}, opposite-sign dilepton \cite{Aaboud:2018jsj,Sirunyan:2019nxl}, and same-sign dilepton and multilepton final states~\cite{Sirunyan:2017uyt,Sirunyan:2017roi,Aaboud:2018xpj}, the combined ATLAS observed (expected) bound on the four-fermion operator \Eq{o4tR} is $|\ctt|/\Lambda^2 < 1.9 \, (1.6) \TeV^{-2}$ at 95\% CL \cite{Aaboud:2018jsj}. A similar bound is obtained by CMS~\cite{Sirunyan:2019nxl}. Besides, very recently both experiments have updated their multilepton searches to $\approx 140$ fb$^{-1}$~\cite{Sirunyan:2019wxt,Aad:2020klt}, observing mild but intriguing excesses with respect to the SM predictions; we will discuss these separately in \Sec{sec:multilep}. Constraints comparable to the one on $c_{tt}$ are obtained for the full set of four-top operators \cite{Sirunyan:2019nxl}, which also involve the third generation left-handed quark doublet,
\beq
\frac{\ctq}{\Lambda^2} (\bar t_R \gamma_\mu t_R) (\bar q_L \gamma^\mu q_L) \, , \,
\frac{\ctqe}{\Lambda^2} \big(\bar t_R \gamma_\mu t^A t_R \big) \big(\bar q_L \gamma^\mu t^A q_L \big) \, , \,
 \frac{\cqq}{\Lambda^2} (\bar q_L \gamma_\mu q_L)^2 \, , \,
 \frac{\cqqe}{\Lambda^2} \big(\bar q_L \gamma_\mu t^A q_L \big)^2 \, ,
\label{o4qL}
\eeq
where $t^A = \lambda^A/2$. In this work our focus is on strongly-interacting right-handed top quarks, for which \Eq{c4tR} sets the size of the associated four-top operator. Then, the generic expectation in scenarios dealing with the generation of the top Yukawa coupling, such as CH models \cite{Panico:2015jxa}, is that operators involving $q_L$ are generated as well, yet with coefficients proportional to $y_t$, i.e.~$\ctq, \ctqe/\Lambda^2 \sim y_t^2/m_*^2$, and $\cqq, \cqqe/\Lambda^2 \sim y_t^2 (y_t/g_*)^2/m_*^2$, thus not as enhanced as $c_{tt}/\Lambda^2$ for large new-physics couplings $g_* \gg y_t$. We note that the $\Hp$ parameter \cite{Englert:2019zmt} effectively contributes to four-top production just as the $O_{tq}$ operator does.

Since our main interest is in new-physics scenarios with a strongly-interacting Higgs, top-Higgs operators should also be present and in principle with large coefficients. This is the case of 
\beq
\frac{\cht}{\Lambda^2}\, i ( H^\dagger \Dlr_\mu H)(\bar t_R \gamma^\mu t_R) \, ,
\eeq
which leads to a zero-momentum deformation of the $Zt_Rt_R$ coupling. However, we point out that, other than the fact that the experimental sensitivity on such anomalous couplings has been typically weak (see however \Sec{sec:multilep}), this operator turns out to be suppressed by an accidental discrete symmetry \cite{Agashe:2006at} in models where the right-handed top does not induce radiative contributions to the Higgs potential, as preferred by fine-tuning considerations. Although such a symmetry is eventually broken, the coefficient of $O_{Ht}$ would be expected to be small in these cases, $c_{Ht}/\Lambda^2 \sim N_c (y_t/4\pi f)^2$.
Similar statements can be made for the analogous operators with $q_L$, namely $O_{Hq}$ and $O_{Hq}^{\text{\tiny (3)}}$ (see \Tab{tab:operators}). The combination $c_{Hq} + c_{Hq}^{\text{\tiny (3)}}$ induces a correction to the $Zb_Lb_L$ coupling which, although constrained at the per-mille level at LEP, is also typically protected by $P_{LR}$ symmetry \cite{Agashe:2006at,Mrazek:2011iu}. Measurements of deviations in the $Zt_Lt_L$ and $Wt_Lb_L$ couplings from the SM, associated with $c_{Hq} - c_{Hq}^{\text{\tiny (3)}}$ and $c_{Hq}^{\text{\tiny (3)}}$ respectively, do not reach the level of precision to be competitive with the four-top operator \Eq{o4tR}, in particular at large $m_*$, because $c_{Hq}/\Lambda^2 \sim y_t^2/m_*^2$. Similarly for the still poor measurements of the Higgs coupling to the top, which probe the Yukawa-like dimension-six operator $O_{y_t}$, even if $c_{y_t}/\Lambda^2 \sim g_*^2/ m_*^2$.

Despite the fact that operators with SM gauge field strengths and top quarks are generated with coefficients that are not enhanced, or are even suppressed at strong coupling,\footnote{Interesting exceptions exist, in particular the so-called Remedios power counting \cite{Liu:2016idz}.}
they could be relevant in situations where direct probes of the four-top operators are not feasible, as it is the case of future lepton colliders (see \Sec{sec:lepton}). 
Such operators are
\beq
\frac{\ctD}{\Lambda^2} (\partial^\mu B_{\mu \nu}) (\bar t_R \gamma^\nu t_R)  \, , \qquad
\frac{\ctDe}{\Lambda^2} (D^\mu G_{\mu \nu}^A) (\bar t_R \gamma^\nu t^A t_R) \, ,
\label{oDtR}
\eeq
with $\ctD/\Lambda^2 \sim g'/m_*^2$ and $\ctDe/\Lambda^2 \sim g_s/m_*^2$, as well as
\beq
\frac{\cqD}{\Lambda^2} (\partial^\mu B_{\mu \nu}) (\bar q_L \gamma^\nu q_L) \, , \quad
\frac{\cqDe}{\Lambda^2} (D^\mu G_{\mu \nu}^A) (\bar q_L \gamma^\nu t^A q_L) \, , \quad
\frac{\cqDt}{\Lambda^2} (D^\mu W_{\mu \nu}^a) (\bar q_L \gamma^\nu \sigma^a q_L), \quad 
\label{oDqL}
\eeq
with $\cqD/\Lambda^2 \sim (y_t/g_*)^2 g'/m_*^2$, $\cqDe/\Lambda^2 \sim (y_t/g_*)^2 g_s/m_*^2$, $\cqDt/\Lambda^2 \sim (y_t/g_*)^2 g/m_*^2$ for a strongly-coupled right-handed top. These operators are equivalent to a particular combination of four-fermion operators, since by the equations of motion,
\begin{align}
\partial^\mu B_{\mu \nu} &= -g' \big( \tfrac{1}{2} i H^\dagger \Dlr_\nu H + \tfrac{2}{3} \bar u_R \gamma_\nu u_R - \tfrac{1}{3} \bar d_R \gamma_\nu d_R + \tfrac{1}{6} \bar q_L \gamma_\nu q_L - \bar e_R \gamma_\nu e_R - \tfrac{1}{2} \bar \ell_L \gamma_\nu \ell_L \big) ,
\label{Beom} \\
D^\mu G_{\mu \nu}^A &= -g_s \big( \bar u_R \gamma_\nu t^A  u_R + \bar d_R \gamma_\nu t^A  d_R + \bar q_L \gamma_\nu t^A  q_L \big),
\label{Geom} \\
D^\mu W_{\mu \nu}^a & = - g \tfrac{1}{2} \big( i H^\dagger \sigma^a \Dlr_\nu H + \bar q_L \gamma_\nu \sigma^a  q_L + \bar \ell_L \gamma_\nu \sigma^a \ell_L \big) \, .
\label{Weom}
\end{align}
For an example of the potential relevance of this class of operators in deciphering the composite nature of the top quark, let us consider probes of $O_{tD}^{\text{\tiny (8)}}$ at the LHC. This operator affects top-pair production through a $q \bar q \to t \bar t$ amplitude that grows with energy \cite{Degrande:2010kt}. Given the expectation $\ctDe/\Lambda^2 \sim g_s/m_*^2$, one could naively conclude that the new-physics effects do not depend on $g_*$ for fixed $m_*$. However, renormalization group evolution implies that at relevant scale, $\mu$, the coefficient of $O_{tD}^{\text{\tiny (8)}}$ is \cite{Lillie:2007hd}
\beq
\ctDe(\mu)= \ctDe(m_*) + \ctt(m_*) \frac{g_s}{12 \pi^2} \log \Big(\frac{m_*^2}{\mu^2}\Big) \,.
\label{rgectDe}
\eeq
Therefore, one-loop diagrams with one insertion of the four-top contact interaction \Eq{o4tR} dominate the amplitude at large $g_*$, $\mathcal{M}_{q \bar q \to t \bar t} \sim g_s^2 (g_*/4\pi)^2 (E/ m_*)^2 \log(m_*^2/E^2)$.
Although current LHC searches in top-pair production yield $\ctDe/\Lambda^2 < 0.7 \TeV^{-2}$ at 95\% CL \cite{Farina:2018lqo} and are therefore not sensitive enough to yield a relevant constraint on the $(m_*,g_*)$ plane, we show in \Sec{sec:lepton} that this changes at high-energy lepton colliders, due to the superior precision in top-pair production.

The main conclusion of the previous discussion is that probes of the four-top operator \Eq{o4tR} are the most relevant ones concerning a strongly-interacting (right-handed) top quark.\footnote{See also~\cite{Kumar:2009vs,Liu:2015hxi} for previous phenomenological studies at the LHC with a similar spirit.} The impact of current LHC bounds from four-top production on the $(m_*,g_*)$ parameter space is shown in \Fig{fig:LHCbounds}, where we also present a comparison with the main universal tests of CH models. 
The latter comprise searches for anomalous Higgs couplings, primarily controlled by $O_H = \tfrac{1}{2} \big( \partial_\mu |H|^2 \big)^2$ and constrained by Higgs and electroweak precision data. The current exclusive (one operator at a time) 95\% CL bound is $c_H/\Lambda^2 < 1.5 \TeV^{-2}$ \cite{Cepeda:2019klc}, which constitutes, given that $c_H/\Lambda^2 \sim g_*^2/m_*^2$, the leading constraint at strong coupling. Note however that such a bound is largely correlated with other contributions to the electroweak parameters, in particular $\Sp$ and $\Tp$, controlled by the operators $O_W$, $O_B$,
and $O_T$ respectively (see \Tab{tab:operators}). This is why the marginalized bound on $O_H$ (i.e.~letting $\Sp$ and $\Tp$ float), determined mainly from LHC data, is significantly weaker, $f|_H^\lhc \gtrsim 550 \GeV$. These last operators are in fact very important in CH models, giving rise to constraints that are independent of the new-physics coupling, since $c_{W,B}/\Lambda^2 \sim 1/m_*^2$ and $c_{T} \sim N_c y_t^2 (y_t/4 \pi)^2/m_*^2$, the latter of one top-loop size because of custodial symmetry \cite{Sikivie:1980hm}. The region of parameter space covered by the bound $(c_W+c_B)/\Lambda^2 < 0.07 \TeV^{-2}$ \cite{Ellis:2018gqa,Cepeda:2019klc}, corresponding to $m_* > 3.7\, \TeV$, is also shown in \Fig{fig:LHCbounds}. 
The other set of relevant bounds are associated with non-standard effects that are largest at weak $g_*$. They are described in terms of the parameters $\Zp$, $\Wp$, $\Yp$ \cite{Barbieri:2004qk}, or equivalently by the operators $O_{2G}$, $O_{2W}$, $O_{2B}$ (see \Tab{tab:operators}), 
with coefficients $c_{2G}/\Lambda^2 \sim (g_s/g_*)^2/m_*^2$, $c_{2W}/\Lambda^2 \sim (g/g_*)^2/m_*^2$, $c_{2B}/\Lambda^2 \sim (g'/g_*)^2/m_*^2$, respectively.\footnote{Since we are considering scenarios with (partial) top quark compositeness, it is implicitly assumed that the new-physics sector features colored states that generate $O_{2G}$ at low energies.}
We find that the current main sensitivity arises from LHC dijet searches, which lead to the bound $c_{2G}/\Lambda^2 < 0.01 \TeV^{-2}$ \cite{Domenech:2012ai,Aaboud:2017yvp,Alioli:2017jdo}. As shown in \Fig{fig:LHCbounds}, this is superior to the LEP and LHC limits derived from $c_{2W}$ and $c_{2B}$ \cite{Cepeda:2019klc}. This remains the case at the high-luminosity phase of the LHC, even if with more statistics the constraint on $\Wp$ from $pp \to \ell \nu$ is expected to reach a comparable level to that on $\Zp$ \cite{Farina:2016rws}. Finally, one of the most relevant constraints on CH models, connected as well with the top sector, comes from the CP-violating operator $\widetilde O_\gamma = H^\dagger H B^{\mu \nu}\widetilde B_{\mu \nu}$, with coefficient of one top-loop size $\tilde c_\gamma/\Lambda^2 \sim g'^2 N_c (y_t/4\pi)^2/m_*^2$ and which itself contributes at one loop to the EDM of the electron. The current constraint on $m_*$, taking the power counting estimate at face value, is $m_* > 20 \TeV$ at the 95\% CL \cite{Panico:2017vlk,Panico:2018hal}. While this is the strongest bound independent of $g_*$, let us note that being CP-violating it is of a qualitatively different nature with respect to the previous ones; for this reason, we do not show it in \Fig{fig:LHCbounds}.

To conclude this section we would like to stress that our approach is markedly driven by a well-motivated yet specific framework. This is why in \Fig{fig:LHCbounds}, as well as in the rest of the paper, the constraints on the $(m_*,g_*)$ parameter space follow from experimental limits on a single operator at a time, each corresponding to the leading new-physics contribution to the relevant observable(s). This viewpoint is therefore orthogonal to that of global top-sector fits, for which the reader is directed to \cite{Buckley:2015nca,AguilarSaavedra:2018nen,Hartland:2019bjb,Durieux:2019rbz,Brivio:2019ius}.

\section{LHC multilepton + jets measurements, and a BSM interpretation}  \label{sec:multilep}
At the LHC, the measurements of $t\bar{t}W$, $t\bar{t}Z$, $t\bar{t}h$, and $t\bar{t}t\bar{t}$ are dominated by final states with multileptons plus jets. After Run 2, an intriguing, generalized pattern of mild excesses has emerged in these final states. We review the latest experimental results here:

\begin{enumerate}
\item  The CMS four-top analysis~\cite{Sirunyan:2019wxt}, in its cut-based version (on which we focus below), finds $\sigma_{t\bar{t}t\bar{t}} = 9.4^{+6.2}_{-5.6} \fb$ compared to the SM reference prediction $\sigmar_{t\bar{t}t\bar{t}} = 12.0 \fb$, while letting the normalizations $\mu_{t\bar{t}W} = 1.3 \pm 0.2$ and $\mu_{t\bar{t}Z} = 1.3 \pm 0.2$ float in the fit, with SM reference cross sections $\sigmar_{t\bar{t}W} = 610 \fb$ and $\sigmar_{t\bar{t}Z} = 840 \fb$, respectively. 
\item The ATLAS four-top measurement~\cite{Aad:2020klt} finds $\sigma_{t\bar{t}t\bar{t}} = 24^{+ 7}_{- 6} \fb$, and also observes an excess of $t\bar{t}W$ events relative to the SM reference, with best fit $\mu_{t\bar{t}W} = 1.6\pm 0.3$ based on $\sigmar_{t\bar{t}W} = 601 \fb$. The normalization of $t\bar{t}Z$ is not left to vary in the fit.
\item The CMS measurement of $t\bar{t}h$, $t\bar{t}W$, and $t\bar{t}Z$~\cite{Sirunyan:2020icl},\footnote{This analysis also measures the $th$ cross section, but we omit it since the accuracy is much weaker than for the $t\bar{t}X$ processes.}
 quotes $\mu_{t\bar{t}W} = 1.43\pm 0.21$ for a reference cross section $\sigmar_{t\bar{t}W} = 650 \fb$. Interestingly, this analysis included for the first time the $O(\alpha_s \alpha^3)$ contribution to $t\bar{t}W+\,$jets at the differential level, dominated by $tW$ scattering~\cite{Dror:2015nkp}. In addition, the fit gives $\mu_{t\bar{t}Z} = 1.03 \pm 0.14$ with $\sigmar_{t\bar{t}Z} = 839 \fb$ and $\mu_{t\bar{t}h} = 0.92^{+0.25}_{-0.23}$ for $\sigmar_{t\bar{t}h} = 507 \fb$.
\item The ATLAS analysis~\cite{ATLAS:2019nvo} finds $\mu_{t\bar{t}W} = 1.39^{+0.17}_{-0.16}$ for a SM reference $\sigmar_{t\bar{t}W} = 727 \fb$ and $\mu_{t\bar{t}h} = 0.70^{+0.36}_{- 0.33}$ for $\sigmar_{t\bar{t}h} = 507 \fb$, when using a single $t\bar{t}W$ normalization factor.
\item A dedicated measurement by CMS of $t\bar{t}Z$~\cite{CMS:2019too} finds a cross section $\sigma_{t\bar{t}Z} = 950 \pm 80 \fb$ compared to the SM reference $\sigmar_{t\bar{t}Z} = 860 \fb$.
\item A dedicated measurement of $t\bar{t}Z$~\cite{ATLAS:2020cxf} by ATLAS obtains $\mu_{t\bar{t}Z} = 1.19 \pm 0.12$ with $\sigmar_{ t\bar{t}Z} = 880 \fb$.
\end{enumerate}
Most measurements~\cite{Sirunyan:2019wxt,Aad:2020klt,Sirunyan:2020icl,ATLAS:2020cxf} are based on $\approx 140\fb^{-1}$, whereas~\cite{ATLAS:2019nvo,CMS:2019too} use $\approx 80 \fb^{-1}$. In addition, we mention but do not discuss further the combined analysis of EFT operators in top associated production modes by CMS~\cite{CMS:2020pnn}, as well as previous measurements of $t\bar{t}W$, $t\bar{t}Z$ by CMS~\cite{Sirunyan:2017uzs} and ATLAS~\cite{Aaboud:2019njj}, which are all based on a smaller data set, $\approx 40 \fb^{-1}$.

\begin{figure}[!t]
    \centering
    \includegraphics[width=0.6\textwidth]{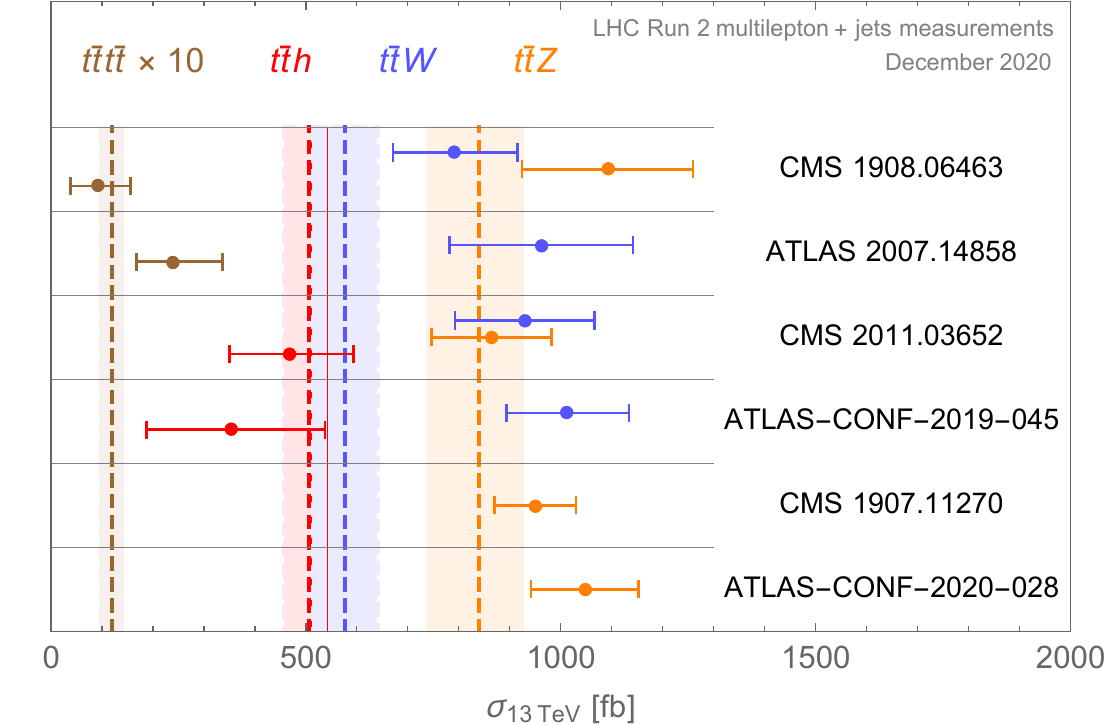}
    \caption{Summary of experimental measurements (dots with error bars) compared to theoretical predictions (dashed vertical lines with uncertainty bands).}
    \label{fig:exp_status}
\end{figure}

The above overview shows that an extensive pattern of $2\sigma$-level excesses with respect to the SM predictions is observed by both ATLAS and CMS. We summarize the status in \Fig{fig:exp_status}, where the above-quoted experimental results are compared with the following SM theoretical predictions in femtobarns
\begin{equation}
\sigma^\sm_{t\bar{t}t\bar{t}} = 12.0^{+18 \%}_{-21 \%}\,, \,\,\,\,
\sigma^\sm_{t\bar{t}W} = 577^{+11\%}_{-11\%} \;\; \text{\cite{Frederix:2017wme}}; \,\,\,\,
\sigma^\sm_{t\bar{t}Z} = 839^{+ 10\%}_{- 12\%}\; , \,\,\,\,
\sigma^\sm_{t\bar{t}h} = 507^{+ 6.8\%}_{- 9.9 \%}\; \;\text{\cite{deFlorian:2016spz}} \,.
\label{thsigma}
\end{equation}
The discrepancies are mild, yet their somewhat coherent structure hints that they may not be due to mere statistical fluctuations.\footnote{Note that 8 TeV data only afforded to measure these processes with order $50\%$ uncertainties; see e.g.~\cite{Khachatryan:2014ewa} for $t\bar{t}W$, $t\bar{t}Z$.} Recent theoretical studies have focused on pushing the SM predictions to higher accuracy, especially for $t\bar{t}W$~\cite{Broggio:2019ewu,Kulesza:2020nfh,Frederix:2020jzp,Bevilacqua:2020pzy,Denner:2020hgg,vonBuddenbrock:2020ter,ATLAS:2020esn}, which however still exhibits the strongest disagreement between theory and experiment. For the time being, a complete NNLO QCD calculation remains out of reach.

Here we take a different point of view and entertain the possibility that the excesses are due to heavy new physics, described by the two effective operators $O_{tt}$ and $O_{Ht}\,$. The former mediates $t\bar{t}t\bar{t}$ production, whereas the latter modifies the $Z t_R t_R$ coupling, thereby leading to three distinct effects: it contributes to $t\bar{t}Z$ production at the leading $O(\alpha_s^2 \alpha)$, to $t\bar{t}W+\,$jets at the formally subleading, but $tW$ scattering-enhanced, $O(\alpha_s \alpha^3)$~\cite{Dror:2015nkp}, and to $t\bar{t}t\bar{t}$ production at $O(\alpha_s^2 \alpha^2)$. We consider these two operators as a motivated first approximation, but note that others should be added in a more general analysis that includes e.g.~$t\bar{t}h$ production, notably $O_{y_t}$ given that this modifies the $htt$ coupling (the sensitivity of $t\bar{t}t\bar{t}$ to $O_{y_t}$ was studied in~\cite{Cao:2016wib}). We concentrate on the CMS four-top analysis~\cite{Sirunyan:2019wxt} because it provides a cut-and-count version and sufficient pre-fit information for us to perform a detailed, if simplified, reinterpretation. 
\begin{figure}[t]
    \centering
    \includegraphics[width=0.48\textwidth]{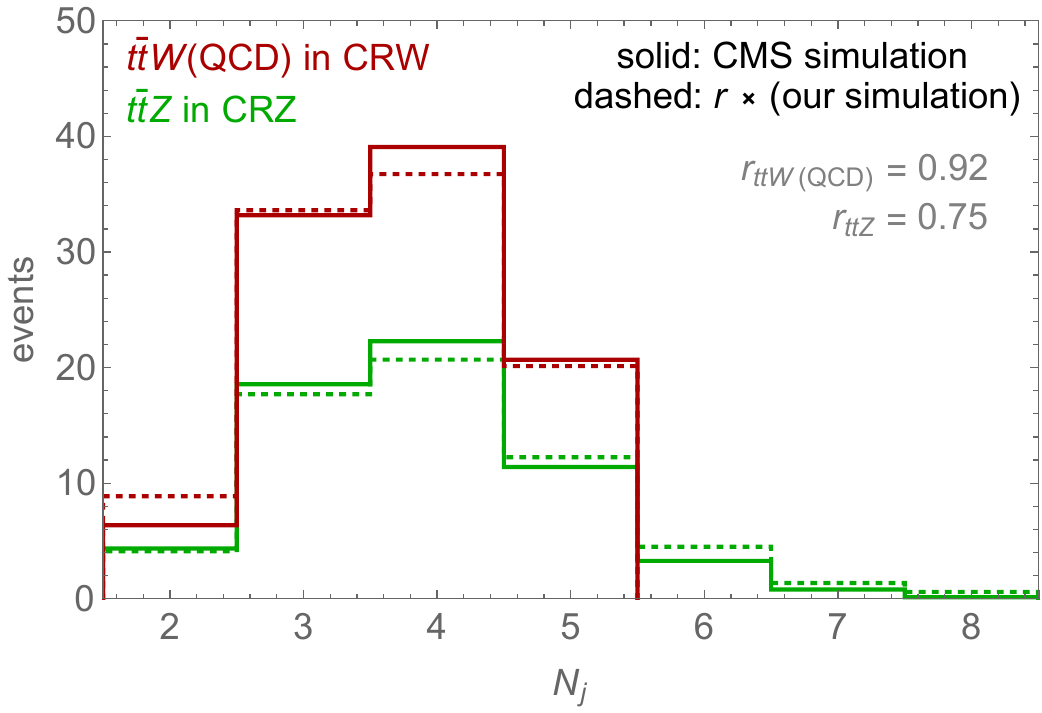}
    \caption{Comparison of our MC event yields to the CMS simulation. Once normalized to $544 \fb$~\cite{Frederix:2017wme}, the $t\bar{t}W$(QCD) sample requires a rescaling factor $r_{t\bar{t}W(\mathrm{QCD})} = 0.92$ to match the total CMS MC yield in CRW. After using $839 \fb$ as normalization~\cite{deFlorian:2016spz}, the $t\bar{t}Z$ sample is rescaled by $r_{t\bar{t}Z} = 0.75$ to match the total CMS MC yield in CRZ.}
    \label{fig:CRW_CRZ}
\end{figure}

The analysis selects events containing at least two leptons of the same sign, $N_j \geq 2$ and $N_b \geq 2$, $H_T > 300$~GeV and $p_T^{\rm miss} > 50$~GeV, with complete definitions and list of requirements reported in~\cite{Sirunyan:2019wxt}. The cut-based analysis defines two control regions, CRW (where the contribution of $t\bar{t}W$ is enhanced) and CRZ (where $t\bar{t}Z$ is enhanced), and $14$ signal regions. In our reinterpretation, for simplicity we combine all signal regions into a single one (SR). Signal and background events are generated using MadGraph5$\char`_\hspace{0.15mm}$aMC@NLO \cite{Alwall:2014hca}, implementing higher-dimensional operators via FeynRules~\cite{Alloul:2013bka}. The factorization and renormalization scales are set to the default dynamical value for all processes, the top mass is set to $172.5$~GeV, and NNPDF23$\char`_\hspace{0.15mm}$lo$\char`_\hspace{0.15mm}$as$\char`_\hspace{0.15mm}$0130$\char`_\hspace{0.15mm}$qed parton distribution functions~\cite{Ball:2013hta} are used. Parton showering and hadronization are performed by Pythia8~\cite{Sjostrand:2014zea} and detector effects are parametrized using the CMS card in Delphes3~\cite{deFavereau:2013fsa}, but setting $R = 0.4$ for the anti-$k_t$ jet clustering algorithm~\cite{Cacciari:2008gp}, implemented via the FastJet package~\cite{Cacciari:2011ma}. As a preliminary check of our simulation tool chain we reproduce the SM $t\bar{t}W$~($t\bar{t}Z$) event yields in the CRW~(CRZ), obtained by CMS with full detector simulation. The results, reported in \Fig{fig:CRW_CRZ}, show that after application of mild scaling factors to match the overall normalizations, our simulations reproduce reasonably well the results reported by CMS (where it should be kept in mind that we simulate at LO in QCD, whereas the CMS treatment is at NLO). Having thus gained confidence in our setup, we proceed to include the new physics effects. 

The impact of $O_{Ht}$ on $t\bar{t}Z$ production is captured by rescaling the CMS yields using the overall factor (note that we set $\Lambda = v$ throughout this section)
\begin{equation}
\mu_{t \bar{t} Z} (\cht) = \frac{g_{Z t_L t_L}^2 + g_{Z t_R t_R}^2 \big(1 +  \frac{3 c_{Ht}}{4 s_w^2} \big)^2}{ g_{Z t_L t_L}^2 + g_{Z t_R t_R}^2 }\,, \qquad g_{Z f f} = g_Z (T_{L f}^3 - s_w^2 Q_f)\,,
\end{equation}
which we have checked to be a good approximation by simulating a set of samples with different values of $\cht$ (see~\cite{Rontsch:2014cca,Bylund:2016phk} for NLO QCD analyses of the $t\bar{t}Z$ sensitivity to top electroweak couplings). In addition, we take into account the impact of $\cht$ on $t\bar{t}Wj$(EW); this piece was altogether neglected in the CMS simulation of $t\bar{t}W+\,$jets~\cite{Sirunyan:2019wxt}. For the $t\bar{t}t\bar{t}$ process, our simulation is simplified in two ways: we neglect interference of the $O_{tt}\,$-$\,$mediated amplitude with the SM,\footnote{The $O(c_{tt}^2)$ term of the cross section is normalized by applying a $K$-factor of $1.24$, as derived for SM four-top production using the NLO QCD-only cross section of $11.1 \fb$~\cite{Frederix:2017wme}. However, very recently the SMEFT@NLO framework~\cite{Degrande:2020evl} has enabled the calculation at full NLO in QCD of the contributions of four-top operators to $t\bar{t}t\bar{t}$ production (including interference with the SM). In particular, $K < 1$ was obtained for the $O(c_{tt}^2)$ piece. Due to the different scale choices, our approximate-NLO cross section turns out to be numerically very close to the exact-NLO result quoted in~\cite{Degrande:2020evl}.} and neglect the contribution of $O_{Ht}$. We do so because reliably assessing these effects at the hadronic differential level goes beyond our computational resources, and besides it would be best performed by the LHC experiments directly. At the qualitative level, we note that the $c_{tt}\,$-$\,$SM interference is suppressed at high energies, whereas the impact of $\cht$ on four-top production is generally expected to be moderate, as the $t\bar{t}\to t\bar{t}$ amplitude does not grow with energy when $\cht \neq 0$, in contrast with the aforementioned case of $tW$ scattering. We provide an estimate of the expected size and pattern of these effects after presenting the results of our fit.

\begin{figure}[t]
	\centering
\raisebox{0\height}{\includegraphics[width=0.4\textwidth]{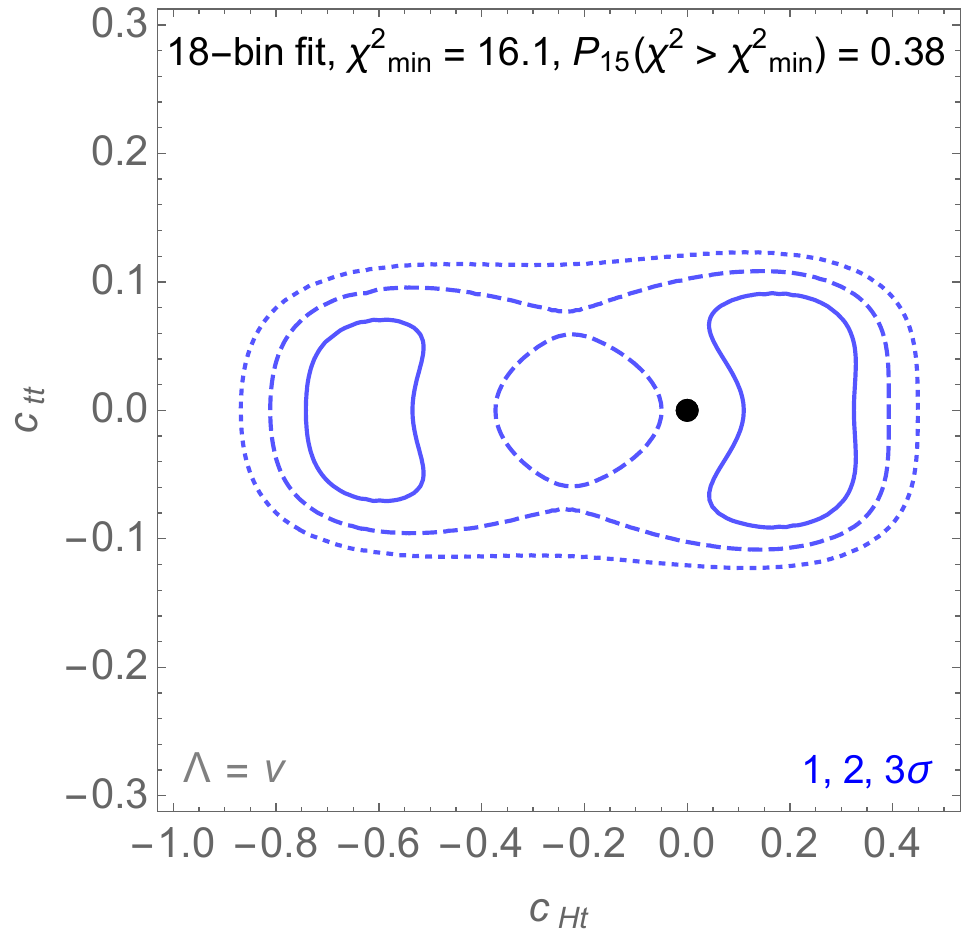}} 
\hspace{2mm}
\raisebox{+0.\height}{\includegraphics[width=0.4\textwidth]{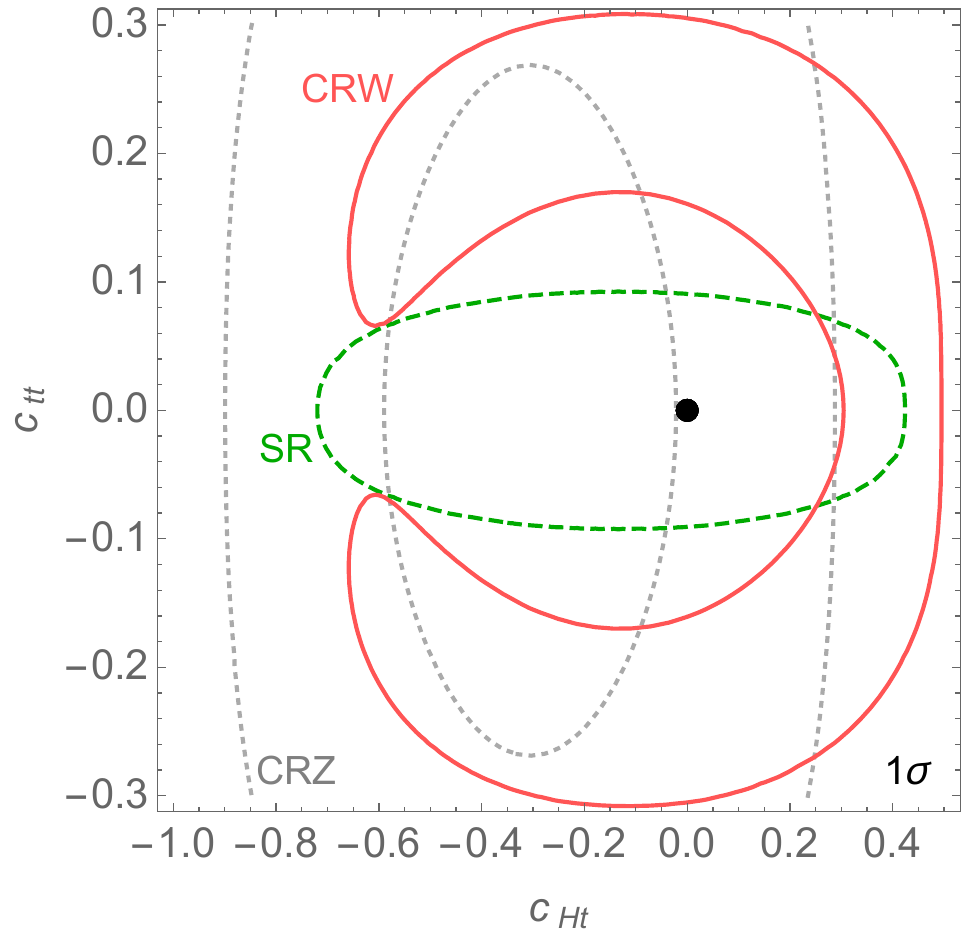}} 
\hspace{0.05cm}
\raisebox{-1\height}{\includegraphics[width=0.4\textwidth]{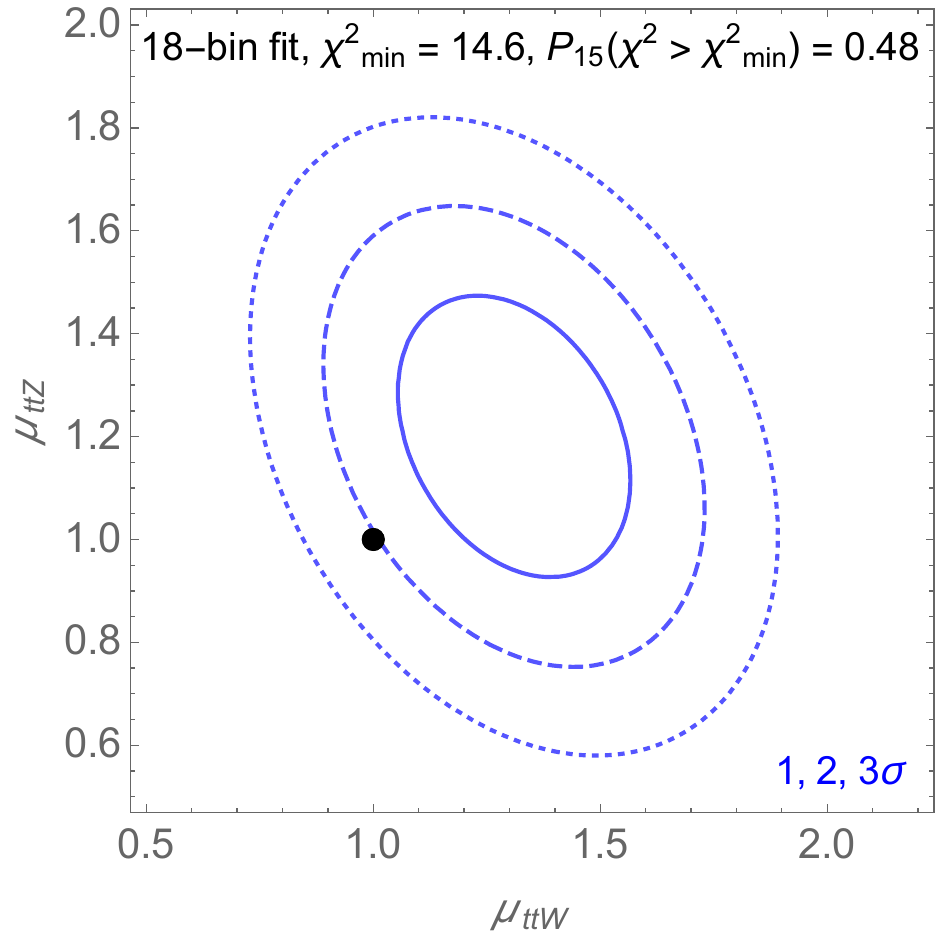}}
	\caption{Exclusion contours from fits to the CMS data in ~\cite{Sirunyan:2019wxt}. {\it Top:} plane of the EFT coefficients $(\cht, \ctt)$. The contours are invariant under $\ctt \to - \ctt$ because we neglected interference between the BSM and SM $t\bar{t}t\bar{t}$ amplitudes. The left panel shows the full fit, whereas the right panel displays the $1\sigma$ regions when the $\chi^2$ is restricted only to CRW, CRZ, or SR. {\it Bottom:} plane of the signal strengths $(\mu_{t \bar{t} W}, \mu_{t \bar{t} Z})$. Here $\mu_{t \bar{t} W}$ rescales the total SM rate, including the $t\bar{t}Wj$(EW) component which we add to the CMS reference cross section. In all panels, a black dot indicates the SM point.}
	\label{fig:fits_LHC}
\end{figure}

\begin{figure}[!t]
	\centering
\raisebox{0\height}{\includegraphics[width=0.45\textwidth]{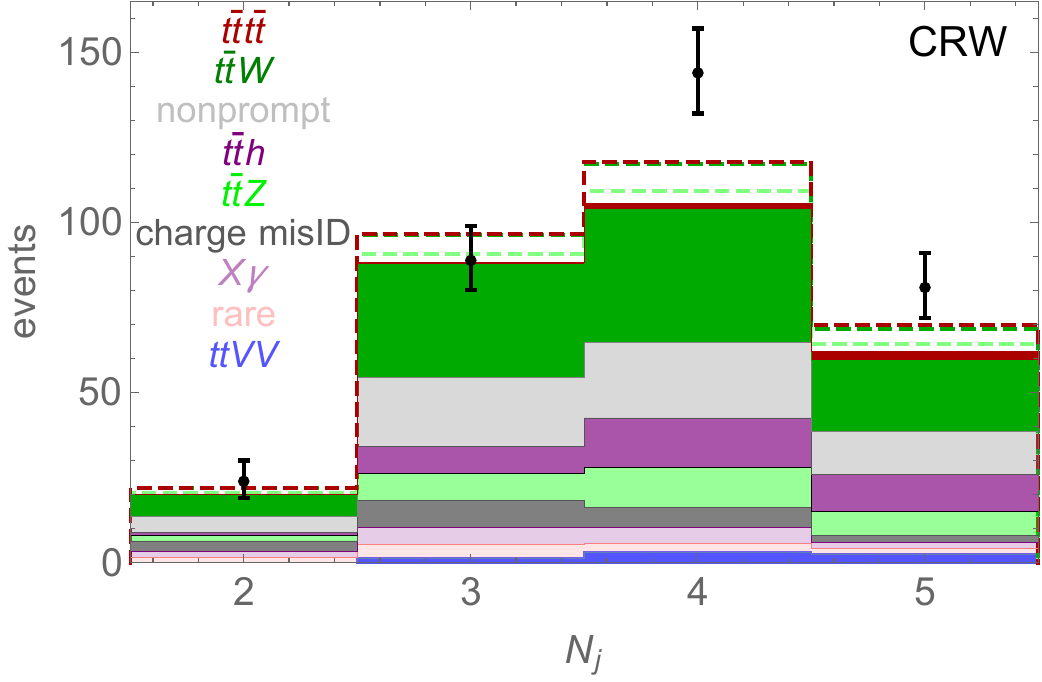}} 
\hspace{2mm}
\raisebox{0.005\height}{\includegraphics[width=0.44\textwidth]{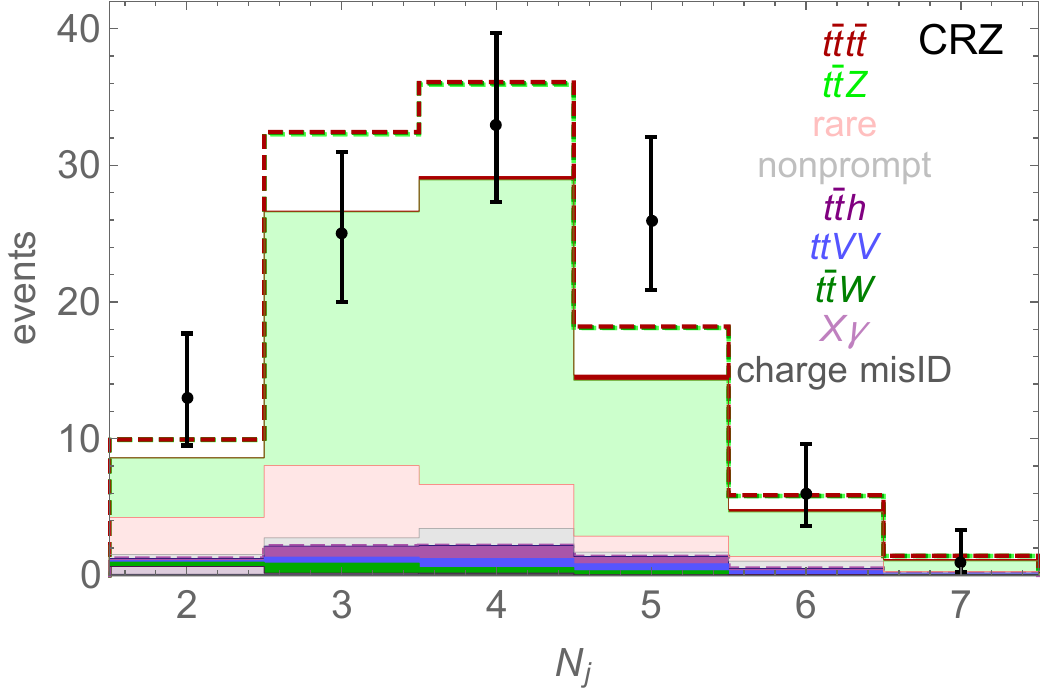}}
\hspace{0.5mm}
\raisebox{-1\height}{\includegraphics[width=0.47\textwidth]{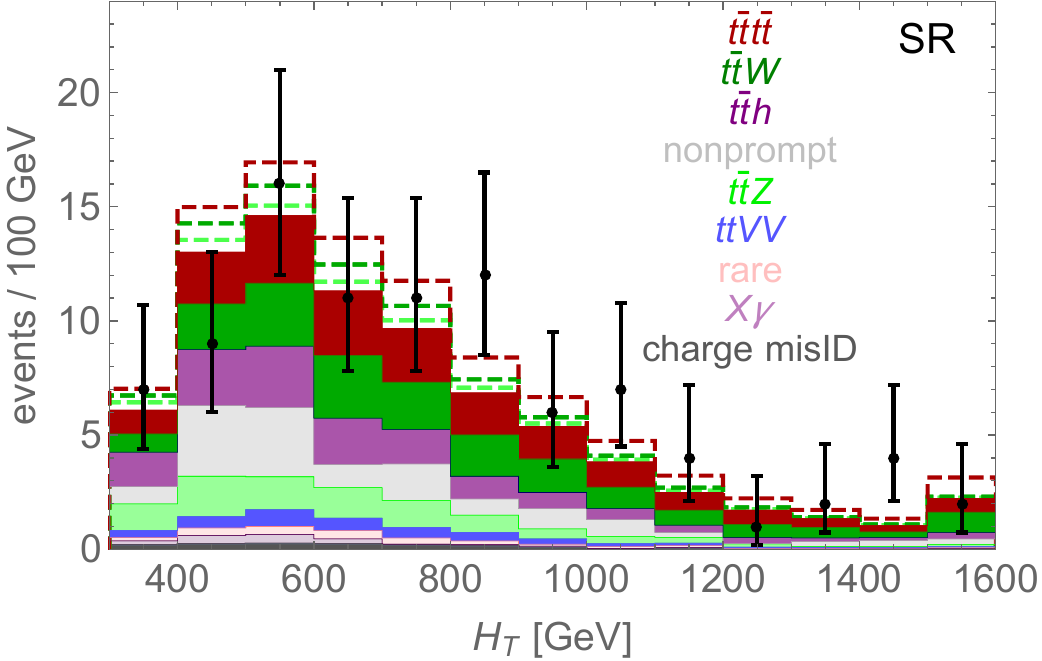}}
	\caption{The filled histograms show the CMS MC predictions as given in~\cite{Sirunyan:2019wxt}. The stacked, light green [dark green]~(dark red) dashed histograms show the BSM contribution to $t\bar{t}Z$~[SM+BSM contribution to $t\bar{t}Wj$(EW)]~(BSM contribution to $t\bar{t}t\bar{t}$) at the best-fit point $(\cht, \ctt) \approx (0.21, \pm 0.054)$. Black points show the data with error bars as quoted by CMS. Although in our fit we combine the last two bins of CRZ and the last 5 bins of SR, for illustration here we retain the same binning as chosen by CMS.}
	\label{fig:histograms_LHC}
\end{figure}

To shed light on the compatibility of the data with our BSM hypothesis, we form a $\chi^2$ from $18$ non-overlapping bins.\footnote{Events are binned by $N_j$ in CRW (4 bins) and CRZ (5 bins, since the $N_j = 6,7$ bins are merged), while in the SR we use 9 $H_T$ bins, where all events with $H_T > 1.1$~TeV are merged into a single bin.} We use the uncertainties on event counts as read from Figs.~2 and 3 in~\cite{Sirunyan:2019wxt}, averaging over positive and negative directions, and neglect theoretical uncertainties. The results of a two-parameter fit to $(\cht, \ctt)$ are shown in the upper panels of \Fig{fig:fits_LHC}, while in the lower panel we show for comparison a fit where no higher-dimensional operators are introduced but the signal strengths $(\mu_{ttW}, \mu_{ttZ})$ are left floating, which is similar to the treatment performed by CMS. The best-fit point of the latter fit is $(\mu_{t \bar{t} W}, \mu_{t \bar{t} Z}) \approx (1.3, 1.2)$. We note that the two EFT coefficients parametrizing the effects of heavy new physics provide a reasonable fit to the data, with comparable goodness of fit to the ad-hoc signal strengths. The best fit is given by $( c_{Ht}, c_{tt} ) \approx (0.21, \pm 0.054)$, corresponding to scales $f|_{Ht} \approx 540 \GeV$ and $f|_{tt} \approx 1.1 \TeV$ if the respective coefficients are set to unity. The impact of the BSM contributions on the CRW, CRZ and SR are shown in \Fig{fig:histograms_LHC} taking the best-fit values of the coefficients.

Next, to gain some insight on the effects of the approximations we made in our description of the four-top process, we consider parton level $t\bar{t}t\bar{t}$ production (with undecayed tops) including the full LO amplitude for the SM plus $O_{tt}$ and $O_{Ht}$. We split the cross section into a low-energy and a high-energy region according to $M_T = \sum_{i = 1}^4 (m_t^2 + p_T^{i\,2})^{1/2}$,
\begin{align} 
\sigma_{M_T\, <\, 1.15 \TeV} \,[\mathrm{fb}] &\,= 6.1 - 20 \, c_{tt} + 410 \, c_{tt}^2 + 5.3 \, c_{Ht} + 9.3 \, c_{Ht}^2 - 63 \, c_{Ht} c_{tt} \,, \label{eq:parton_xsec1} \\
\sigma_{M_T \,>\, 1.15 \TeV} \,[\mathrm{fb}] &\,= 3.6 - 6.8\, c_{tt} + 1100\, c_{tt}^2 + 1.4\, c_{Ht} + 2.3\, c_{Ht}^2 - 25\, c_{Ht} c_{tt} \,. \label{eq:parton_xsec2}
\end{align}
The boundary value $M_T = 1.15 \TeV$ is chosen to roughly match $H_T = 800 \GeV$ at hadronic level, which we have verified splits the SR into two sub-regions of comparable sensitivity in our fit to CMS data (see the bottom panel of \Fig{fig:histograms_LHC}). Equation~(\ref{eq:parton_xsec2}) confirms the expectation that at high energies, it is reasonable to neglect all BSM terms except for the $O(c_{tt}^2)$ one: for example, plugging in the best fit point, we find $\sigma_{>} / \sigma_{>}^\sm = 1.79$ whereas our approximation gives $1.86$. For the low-energy region, using \Eq{eq:parton_xsec1} we find $\sigma_{<} / \sigma_{<}^\sm = 1.15$ versus the approximate value $1.20$. This apparently reasonable agreement is, however, actually the result of a compensation between different corrections arising from $\cht$ and $\ctt$, suggesting that the shapes of our fit contours could be somewhat affected by a fully accurate description of BSM effects in the low-$H_T$ bins of SR.

Finally, we remark that $O_{Ht}$ mediates BSM contributions to additional processes, including for instance $pp\to t\bar{t}hj$ at $O(\alpha_s \alpha^3)$ and $tZW$ at $O(\alpha_s \alpha^2)$. The analysis of such subleading effects was initiated in~\cite{Dror:2015nkp} and later expanded in~\cite{Maltoni:2019aot}. Based on their findings we do not expect the $O_{Ht}$ dependence of these and other analogous processes, which is neglected here, to have a significant impact on our results. Nonetheless, a detailed assessment would be of interest to obtain a complete picture of heavy new physics effects in LHC multilepton plus jets final states.

In summary, the main messages we derive from the fit are:
\begin{itemize}
\item The $O(\alpha_s \alpha^3)$ $t\bar{t}Wj$(EW) contribution to $t\bar{t}W+\,$jets is important and should be consistently included at the differential level, as originally pointed out in~\cite{Dror:2015nkp} and later analyzed in depth in~\cite{Frederix:2017wme,Frederix:2020jzp}. 
\item An interpretation of the CMS data~\cite{Sirunyan:2019wxt} in terms of the $O_{Ht}$ and $O_{tt}$ operators gives a goodness of fit comparable to the application of constant rescaling factors to the $t\bar{t}W$ and $t\bar{t}Z$ cross sections, while having a stronger physical motivation.
\item While it is too early to draw any conclusions, it is intriguing that a scale $f \sim 750 \GeV$ improves the fit to multilepton + jets data, while being roughly consistent both with four-top constraints from the single lepton and opposite-sign dilepton final states~\cite{Aaboud:2018jsj}, and with measurements of the Higgs couplings~\cite{Cepeda:2019klc}.
\end{itemize}
A more comprehensive study, including a wider set of signal regions, would be strongly desirable to obtain further insight. Nevertheless, we regard the coincidence of scales suggested by our analysis as an additional motivation to further investigate heavy top-philic new physics.


\section{Future proton-proton colliders}\label{sec:hadron}

Let us remark that under well-motivated assumptions, current searches for strong $t \bar t t \bar t$ production enjoy a higher reach on the compositeness scale $4 \pi f$ than probes of the Higgs sector at the LHC. This fact motivates our sensitivity studies at future colliders. We begin in this section with hadron colliders, first discussing shortly the high-luminosity phase of the LHC and then analyzing in detail the 100 TeV FCC-hh \cite{Abada:2019lih}.

To estimate the HL-LHC sensitivity to $c_{tt}/\Lambda^2$ we perform a simple extrapolation of the CMS four-top search in multileptons~\cite{Sirunyan:2019wxt}. We focus on the signal region (bottom panel of Fig.~\ref{fig:histograms_LHC}), adopting the $H_T$-binning chosen by CMS and rescaling their MC predictions for all SM processes to a luminosity of $3$~ab$^{-1}$. As in the previous section, we include the missing SM $t\bar{t}Wj(\mathrm{EW})$ contribution among the backgrounds and simulate the signal neglecting interference with SM four-top production. Assuming as systematics on the two main backgrounds $(\delta_{t\bar{t}t\bar{t}}, \delta_{t\bar{t}W}) = (8.5\%, 7.5\%)$, which correspond to half the current theoretical uncertainties~\cite{Frederix:2017wme}, and applying a mild PDF rescaling factor~\cite{ColliderReach} to account for the increase of collider energy to $14$~TeV, we obtain at $95\%$~CL
\begin{equation} \label{eq:HL_LHC_bound} 
\qquad\qquad\qquad \Lambda / \sqrt{|\ctt| } \,>\, 1.3 \TeV\;\,(\mathrm{no}\;\mathrm{syst.\hspace{-1mm}:}\;1.4 \TeV) \,. \qquad \qquad \qquad (\mathrm{HL}\,\mbox{-}\,\mathrm{LHC})
\end{equation}
We view this as a conservative estimate, as the actual HL-LHC analysis will capitalize on the $\approx 20$ times larger statistics by refining the binning at larger $H_T$, thus increasing slightly the sensitivity.\footnote{In addition, rescaling the current statistical-only $95\%$ CL bound $\Lambda / \sqrt{ |c_{tt}| } > 0.93$~TeV using Collider Reach~\cite{ColliderReach} would give an estimate of $1.7$~TeV at the HL-LHC.} Furthermore, a caveat is that we have assumed agreement of data with the SM predictions, although as discussed in Sec.~\ref{sec:multilep} this is somewhat unclear for current multilepton measurements.

We now turn to the analysis of the four-top final state at the FCC-hh. The decays of the four tops give rise to a complex set of possible final states. The same-sign dileptons (SSL) and trileptons (3L) signatures both benefit from suppressed SM backgrounds, while retaining not-too-small branching ratios of $4.1\%$ and $2.6\%$, respectively. These numbers do not include contributions from leptonic $\tau$ decays, which are systematically neglected in our FCC analysis (whereas they are always included when we quote LHC results).\footnote{As taus dominantly originate from $W$ and $Z$ decays, they give approximate equal contributions to both signal and backgrounds, hence neglecting them makes our FCC results slightly conservative.} Conversely, the fully hadronic signature has a large branching ratio of $20\%$, but suffers from challenging backgrounds. In this work we thoroughly analyze the SSL and 3L signatures, and perform an exploratory study of the fully hadronic final state.

For the SSL and 3L final states we partly build on the latest LHC searches for four-top production in multilepton$\,$+$\,$jets~\cite{Sirunyan:2019wxt,Aad:2020klt}, and on the LHC study~\cite{Alvarez:2016nrz} which focused on SM $t\bar{t}t\bar{t}$ production and BSM effects mediated by relatively light new physics (see also~\cite{Kim:2016plm} for a thorough analysis of resonant signals in the four-top final state at the LHC). Signal and background events are generated using MadGraph5$\char`_\hspace{0.15mm}$aMC@NLO \cite{Alwall:2014hca}, using a FeynRules~\cite{Alloul:2013bka} model where $O_{tt}$ is added to the SM. The factorization and renormalization scales are set to $M_T/2$ for all processes, where $M_T$ is the sum of transverse masses. The signal samples only contain the $O(c_{tt}^2)$ contribution, as interference with the SM $t\bar{t}t\bar{t}$ amplitude is a small effect in our signal region; we provide a quantitative assessment of this at the end of the section. The SM four-top production is simulated at full LO, namely $O(\alpha_s^i \alpha^j)$ with $i, j \in \{0, \ldots, 4\}$ and $i + j =4$, while as normalization we use the complete NLO (QCD+EW) calculation of~\cite{Frederix:2017wme}. The normalization of the signal is rescaled by the ratio of the NLO (QCD+EW) and LO QCD cross sections as calculated for SM production, which equals $1.8$.

Parton showering and hadronization are performed by Pythia8~\cite{Sjostrand:2014zea} and detector effects are parametrized using Delphes3~\cite{deFavereau:2013fsa} adopting the FCC card. Within Delphes, jets are clustered with the FastJet package~\cite{Cacciari:2011ma} using the anti-$k_t$ algorithm~\cite{Cacciari:2008gp} with $R = 0.5$. The $b$-tagging performance is described through the following efficiencies,
\begin{align} \label{eq:btagging}
\epsilon_i (p_T) &= \epsilon_i^0\, \chi_{[10 \GeV, \,15 \TeV]} (p_T)  \Big( 1 - \chi_{[500 \GeV, \,15 \TeV]}(p_T)\, \frac{p_T}{15 \TeV} \Big)\,,  \\
\epsilon_{b,c,j}^0 = 0.85, 0.05, &0.01 \,\,\, \mathrm{for} \,\,\, |\eta| < 2.5\,, \qquad \epsilon_{b,c,j}^0 = 0.64, 0.03, 0.0075 \,\,\, \mathrm{for} \,\,\, 2.5 < |\eta| < 4.0\,,  \nonumber
\end{align}
and $\epsilon_{b,c,j}^0 = 0$ for $|\eta| > 4.0$. In Eq.~\eqref{eq:btagging}, $\chi$ denotes the characteristic function. As our signals feature highly boosted tops, as well as a generally large amount of hadronic activity, we apply lepton isolation using a variable cone, following the mini-isolation proposal~\cite{Rehermann:2010vq}: an electron~(muon) $\ell$ is said to be isolated if $p_T^{\rm cone} / p^\ell_{T} < 0.1\,(0.2)$, where $p_T^{\rm cone}$ is the sum of the transverse momenta within a cone of radius $R_{\rm iso} = \mathrm{min}\, (r_{\rm min}, p_{T}^0/p^\ell_{T})$ around the lepton [the sum excludes the lepton itself], where $r_{\rm min} = 0.2~(0.3)$ and $p_{T}^0 = 8\,(10) \GeV$. These values are very similar to those used in~\cite{Alvarez:2016nrz,Aaboud:2018jsj}. As input parameters we take
\begin{equation}
G_F = 1.166 \times 10^{-5} \GeV^{-2}, \quad m_{Z,h,t} = 91.19,\, 125,\, 173 \, \GeV, \quad \alpha (m_Z) = 1/132.5\,,
\end{equation}
and we employ NNPDF23$\char`_\hspace{0.15mm}$lo$\char`_\hspace{0.15mm}$as$\char`_\hspace{0.15mm}$0130$\char`_\hspace{0.15mm}$qed parton distribution functions~\cite{Ball:2013hta}.

\subsection{Same-sign dileptons} \label{sec:SSL}
In this channel, the main background beyond the irreducible SM $t \bar{t} t \bar{t}$ is the production of $t\bar{t}W+\,$jets, which is in fact also primarily measured in the SSL final state. Secondary backgrounds with genuine SSL include $t\bar{t}Z$ and $t\bar{t}h$, as well as some other processes listed in Table~\ref{tab:SSL_bkg}, together with the MC generation-level cross sections. In all cases we generate processes giving rise to at least a SSL pair and four jets at the matrix element level; for a few backgrounds, we are able to include additional jets within computing limitations. Some important processes, including SM $t\bar{t}t\bar{t}$ and $t\bar{t}W$ production, are normalized to the best available predictions that include both QCD and EW corrections~\cite{Frederix:2017wme,Contino:2016spe,Torrielli:2014rqa}. 

In addition, there are important reducible backgrounds: either a jet is mis-identified as a ``fake'' lepton, or one lepton belonging to an opposite-sign pair has its charge mismeasured ($Q$flip); both of these originate mainly from $t\bar{t}\,+\,$jets. The fake lepton component can be estimated by applying a probability for a given jet to be misidentified as a lepton (in general, the probability depends on the jet flavor and $p_T$), and a transfer function relating the properties of the daughter lepton to those of the parent jet~\cite{Curtin:2013zua}. The probability and transfer function parameters need to be tuned against data. Here we follow a simplified approach, assuming a constant probability for both heavy flavor and light jets, and that the fake lepton inherits the full four-momentum of the jet it originates from, whereas the lepton charge and flavor are assigned randomly and independently. The probability is fixed to $\varepsilon_{\rm fake} = 3.7 \times 10^{-5}$ by comparing a sample of $13 \TeV$ semileptonic $t\bar{t}\,+\,$jets, normalized to a cross section of $832 \pb$~\cite{Czakon:2011xx}, to the ``nonprompt'' yields in the control region CRW of the CMS four-top search~\cite{Sirunyan:2019wxt} (see left panel of \Fig{fig:histograms_LHC}).\footnote{For comparison, $\varepsilon_{\rm fake} = 7.2 \times 10^{-5}$ was obtained in~\cite{Alvarez:2016nrz} by matching to an earlier ATLAS analysis.} 

The $Q$flip component is estimated from MC events containing an $e^+e^-$ or $e^\pm \mu^\mp$ pair and applying a constant probability for the charge of each electron with $p_T^e > 10 \GeV$ to be mismeasured (the probability of flipping the charge of a muon is negligible). The probability $\varepsilon_{\rm flip} = 2.2 \times 10^{-4}$ is taken from~\cite{Alvarez:2016nrz} and further validated by checking that a $13 \TeV$ fully leptonic $t\bar{t}\,+\,$jets sample reproduces the ``charge misID'' yields in the control region CRW of~\cite{Sirunyan:2019wxt}. The processes we include in our estimates of the fake lepton and $Q$flip backgrounds at the FCC are listed in Table~\ref{tab:SSL_bkg}.
\begin{table}[t]
	\begin{center}
	\begin{tabular}{c|c|c|c}
		category&processes & decay channel & $\sigma \times \mathrm{BR}$ [fb]\\\hline
		\multirow{2}{*}{$t\bar{t}t\bar{t}$ (signal)} & $t\bar{t}t\bar{t}$ & \multirow{2}{*}{$W_{\ell^{\pm}}\:W_{\ell^{\pm}}\:W_{\rm had}\: W_{\rm had}$} & \multirow{2}{*}{$0.325$}\\
		& $\Lambda/\sqrt{| c_{tt}| } = 6 \TeV$ &  & \\\hline
		$t\bar{t}t\bar{t}$ (SM)&$t\bar{t}t\bar{t}$&$W_{\ell^{\pm}}\:W_{\ell^{\pm}}\:W_{\rm had}\: W_{\rm had}$&$144$~\cite{Frederix:2017wme}\\\hline
		\multirow{3}{*}{$t\bar{t}W$} & $t\bar{t}W^{\pm}\,+\,$0,1,2 jets &$W_{\ell^{\pm}}\:W_{\ell^{\pm}}\:W_{\rm had}$&640~\cite{Frederix:2017wme}\\
		&$t\bar{t}W^{\pm}bjj$&$W_{\ell^{\pm}}\:W_{\ell^{\pm}}\:W_{\rm had}$&4.11\\
		&$t\bar{t}W^{\pm}jj$ &$W_{\ell^{+}}\:W_{\ell^{-}}\:W_{\ell^{\pm}}$&63.4$^\dagger$\hspace{0.25mm}\cite{Torrielli:2014rqa}\\\hline
		\multirow{2}{*}{$t\bar{t}Z$} & $t\bar{t}Z\,+\,$0,1,2 jets &$W_{\ell^{\pm}}\:W_{\rm had}\:Z_{\ell^+ \ell^-}$&1120~\cite{Contino:2016spe}\\
		&$t\bar{t}Zjj$ &$W_{\ell^{+}}\:W_{\ell^{-}}\:Z_{\ell^+ \ell^-}$&82.6\\\hline
		\multirow{3}{*}{$t\bar{t}h$} &$t\bar{t}h,\: h\rightarrow WW^*$&$W_{\ell^{\pm}}\:W_{\ell^{\pm}}\:W_{\rm had}\:W_{\rm had}$&300~\cite{Contino:2016spe}\\
		&$t\bar{t}h,\: h\rightarrow ZZ^*$&$W_{\ell^{\pm}}\:W_{\rm had}\:Z_{\ell^+ \ell^-}\:Z_{\rm had}$&24.0~\cite{Contino:2016spe}\\
		&$t\bar{t}h,\: h\rightarrow \tau^+\tau^-$&$W_{\ell^{\pm}}\:W_{\rm had}\:\tau_{\ell^{\pm}}\:\tau_{\rm had}$&140~\cite{Contino:2016spe}\\\hline
		\multirow{3}{*}{other} &$tZbjj$&$W_{\ell^{\pm}}\:Z_{\ell^+ \ell^-}$&145\\
		&$t\bar{t}W^+W^-$&$W_{\ell^{\pm}}\:W_{\ell^{\pm}}\:W_{\rm had}\:W_{\rm had}$&35.3\\
		&$t\bar{t}W^+W^-$&$W_{\ell^{+}}\:W_{\ell^{-}}\:W_{\ell^\pm}\:W_{\rm had}$&11.7\\\hline
		\multirow{2}{*}{fake $\ell$}& $t\bar{t}\,+\,$1,2 jets&$W_{\ell^{\pm}}\:W_{\rm had}$&$K_{t\bar{t}} \;3.45 \times10^6$\\
		&$t\bar{t}bjj$&$W_{\ell^{\pm}}\:W_{\rm had}$&$K_{t\bar{t}} \;6.13 \times 10^{4}$\\\hline
		\multirow{2}{*}{$Q$flip} &$t\bar{t}jj$&$W_{\ell^{+}}\:W_{\ell^{-}}$&$K_{t\bar{t}} \;4.63 \times 10^5$\\
		&$t\bar{t}bjj$&$W_{\ell^{+}}\:W_{\ell^{-}}$&$K_{t\bar{t}} \;1.06 \times 10^4$\\
	\end{tabular} 
	\end{center}
	\caption{SSL signal and background processes at $\sqrt{s}=100 \TeV$. Samples with different jet multiplicities have been merged using the MLM prescription with matching scale of $30 \GeV$. The cuts $p_T^j > 50 \GeV$ and $|\eta_j| < 5$ are imposed on jets arising from QCD radiation, but no cuts are applied yet to decay products of heavy particles. The subsequent baseline selection, discussed in Sec.~\ref{sec:SSL}, requires $\geq 5$ jets among which $\geq 3$ are $b\hspace{0.2mm}$-tagged. The higher-order cross sections we use for normalization always assume $\mu = M_T/2\,$ (note that in~\cite{Frederix:2017wme} this is not the central choice for $t\bar{t}t\bar{t}$). The $^\dagger$ indicates that $p_T^j > 100 \GeV$ was exceptionally required, to match~\cite{Torrielli:2014rqa} (we have checked that this different initial cut has negligible impact on the event yield after the complete selection). Whenever they do not appear in $t\bar{t}$ or $b\bar{b}$ pairs, the symbols $t$ and $b$ refer to either particles or antiparticles. To the $t\bar{t}\,+\,$jets samples used to estimate the fake lepton and $Q$flip backgrounds we apply a $K_{t\bar{t}} = 1.4$, calculated for inclusive $t\bar{t}$ production using the NNLO cross section of $34.7 \nb$~\cite{Mangano:2016jyj}.}
	\label{tab:SSL_bkg}
\end{table}

\begin{figure}[!t]
    \centering
         \raisebox{0\height}{\includegraphics[width=0.495\textwidth]{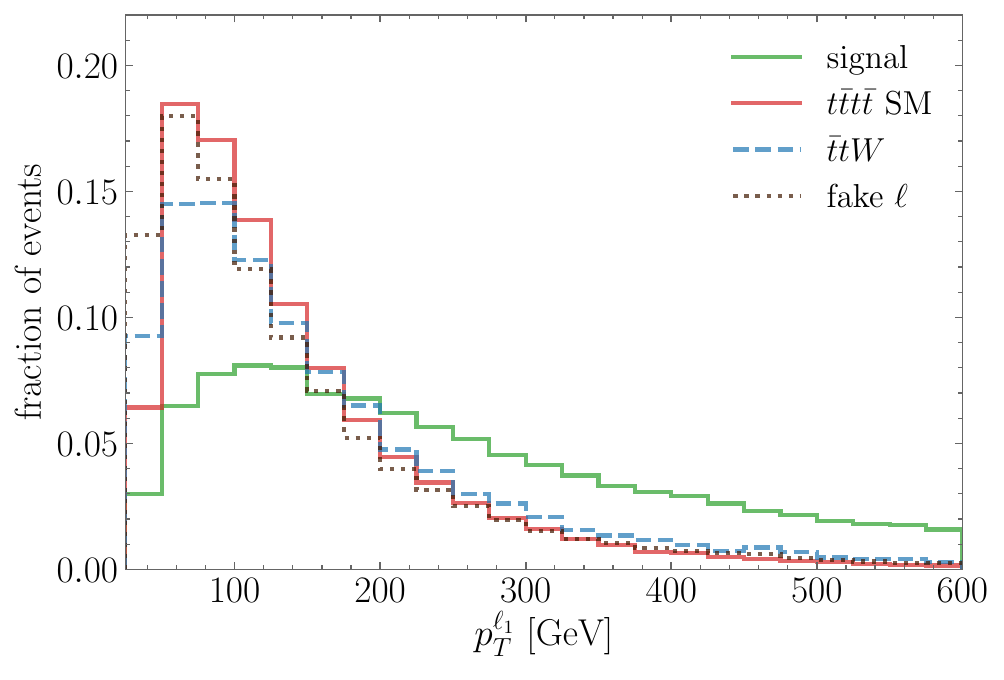}}
     \raisebox{0.009\height}{\includegraphics[width=0.489\textwidth]{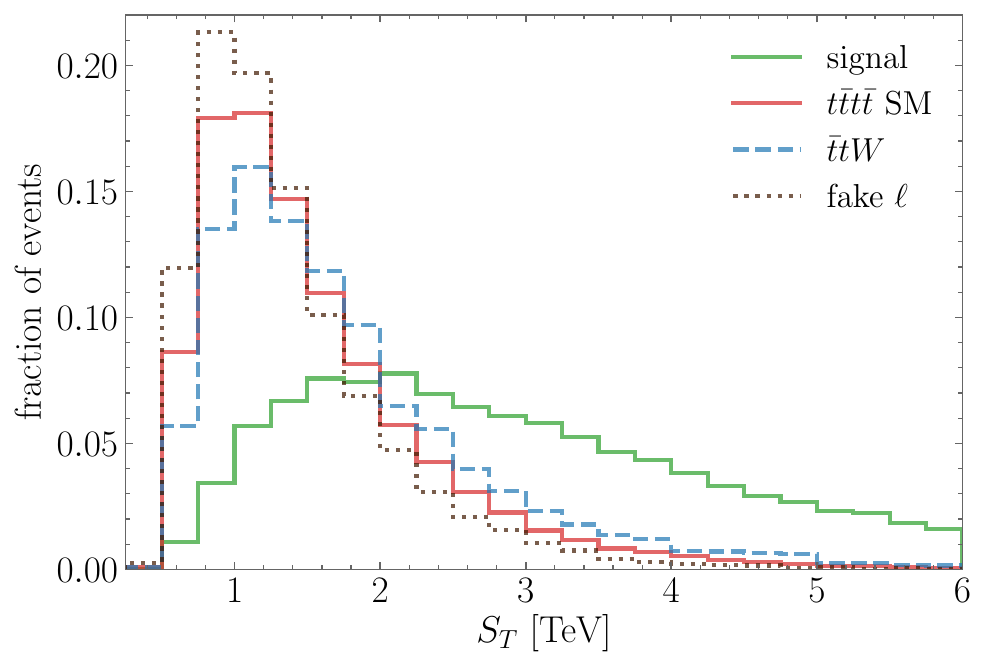}}
    \caption{Normalized distributions of the leading lepton $p_T$ {(left)} and the scalar sum of the transverse momenta of all jets and the two leptons {(right)}, after the baseline SSL selection, for the signal and the main backgrounds.}
    \label{fig:SSL_distrib}
\end{figure}

We now turn to the event selection. First, we identify the lepton and jet candidates satisfying
\begin{equation} \label{eq:candidates}
p_T^\ell > 25\GeV \,, \; |\eta_{\ell}| < 3 \,, \qquad\quad p_T^j > 50\GeV \,, \; |\eta_j| < 5\,.
\end{equation}
Next, to prevent assignment of a single detector response to both a lepton and a jet, we apply to the selected candidates an overlap removal procedure, following closely~\cite{Aaboud:2018jsj}. To avoid the double counting of energy deposits as electrons and jets, for each electron the closest jet within $\Delta R < 0.2$ (if any) is removed; however, if the next-to-closest jet is within $\Delta R < 0.5$ of the electron, then the electron is removed and the previously removed jet is reinstated. For muons we apply a different criterion, aimed at distinguishing muons arising from hadron decays within proper jets, from muons that undergo bremsstrahlung radiation inside the calorimeter and are accidentally reconstructed as jets, typically characterized by a very small number of matching tracks. If a jet satisfies $\Delta R(\mu, j) < 0.04 + 10\GeV/p_T^\mu$ and it has at least three tracks, the muon is rejected; otherwise, the jet is removed. The baseline selection is completed by the following requirements,
\begin{align}
&\qquad\quad\mathrm{exactly\; two\; SSL\;with\;} p_T^{\ell_1, \ell_2} > 40, 25\GeV \,, \nonumber \\ 
&\geq 5 \;\mathrm{jets}, \; \mathrm{of\;which}\; \geq 3\; b\hspace{0.2mm}\mbox{-}\mathrm{tagged} \,, \qquad H_T > 400\GeV \,,
\end{align}
where $\ell_1\,(\ell_2)$ denotes the (sub)leading lepton. We expect that the above requirements on the lepton transverse momenta will allow for a high efficiency of an FCC-hh dilepton trigger. At this stage, for a reference BSM scale $\Lambda/ \sqrt{|\ctt|} = 6 \TeV$, we have $S/B \sim 10^{-3}$, as shown in Table~\ref{tab:SSLcut}. Therefore we search for additional cuts tailored to the signal, which is characterized by a hard $t\bar{t} \to t\bar{t}$ scattering. We find as optimal variables the $p_T$ of the leading lepton and $S_T$, defined as the scalar sum of the transverse momenta of the SSL pair and of all jets. Normalized distributions of these variables after the baseline selection are shown for the signal and the main backgrounds in \Fig{fig:SSL_distrib}. 
\renewcommand{\tabcolsep}{2.7pt}
\begin{table}[t]
	\begin{center}
	\begin{tabular}{c|c|c|c|c|c|c|c|c|c|c}
		& $\mathrm{signal}$ & $t\bar{t}t\bar{t}$ & \multirow{2}{*}{$t\bar{t}W$} & \multirow{2}{*}{$t\bar{t}Z$}&\multirow{2}{*}{$t\bar{t}h$}&\multirow{2}{*}{other} & fake & \multirow{2}{*}{$Q$flip} & $\mathcal{S}\;\mathrm{at}$ &  $S/B$ \\
		& $\Lambda / \sqrt{|\ctt| } = 6 \TeV$ & SM &&&& & $\ell$ & & $30/$ab & $[10^{-2}]$  \\\hline
		baseline&  43 &$ 17000$ & $4200$ & $2900$ & $1800$  & 920 & $5300$ & 2200 & 1.3 & 0.13  \\
		$p_T^{\ell_1 } > 275 \GeV$ & 20 & $1600$ & $670$ & $300 $ & $110$ & 120 & 590 & 130 & 1.8 & 0.55 \\
		$S_T \in [3,4] \TeV$ & 4.2  & $260$  & 120 &  90 & 11 & 13 & 47 & 13 & 0.99 & 0.77  \\
		$S_T \in [4,5] \TeV$ & 3.1  & $110$  & 56 &  1.0 & 5.4 & 6.0 & 15 & 4.1 & 1.2 & 1.6  \\
		$S_T > 5 \TeV$ & 6.1  & $67$  & 41 &  2.1 & 2.5 & 2.6 & 7.3 & 2.4 & 2.9 & 4.9  \\
	\end{tabular}
	\end{center}
	\caption{Cut flow for the SSL final state, with cross sections in ab. The (purely statistical) significance is defined as $\mathcal{S} = S/ \sqrt{S + B}\,$, and a two-sided exclusion at $(1-p)$ CL corresponds to $\mathcal{S} = \sqrt{2}\, \mathrm{erf}^{-1} (1-p)$.}
	\label{tab:SSLcut}
\end{table}
\hspace{-2mm}We apply the cuts $p_T^{\ell_1} > 275 \GeV$ and $S_T > 3 \TeV$, and divide the remaining events into three $S_T$ bins, with $S/B$ ranging from $8\times 10^{-3}$ to $5\times 10^{-2}$. To derive a bound on $\Lambda$ we construct a $\chi^2$, accounting for the systematic uncertainties on the two main SM backgrounds, namely $t\bar{t}t\bar{t}$ and $t\bar{t}W$. For $n$ bins, the $\chi^2$ is defined as
\begin{equation}
\chi^2 = \sum_{i,j = 1}^{n} N^S_{i} (C_{\rm tot}^{-1})_{ij} N^S_{j}\,,\qquad C_{\rm tot} = C_{\rm stat} + C_{\rm sys}^{t\bar{t}t\bar{t}} + C_{\rm sys}^{t\bar{t}W}\,,
\end{equation}
where the number of signal events $N_i^S \propto c_{tt}^2\, \Lambda^{-4}$, and
\begin{equation}
(C_{\rm stat})_{ij} = (\sigma^{\rm stat}_i)^2 \delta_{ij} \,, \qquad (C_{\rm sys}^{A})_{ij} = \sigma^{{\rm sys,} A}_i \rho_{ij} \sigma^{{\rm sys,} A}_j  \qquad (A = t\bar{t}t\bar{t}, \, t\bar{t}W) \,,
\end{equation}
with $\sigma^{\rm stat}_i = \sqrt{N^\sm_i}$, $\sigma^{{\rm sys,} A}_i = \delta_A N_i^{A}$ and we assume that each systematic uncertainty $\delta_A$ is fully correlated across bins, namely $\rho_{ij} = 1$ for all $i,j$. We have also assumed that the observed number of events will match the SM expectation. We take $(\delta_{t\bar{t}t\bar{t}}, \delta_{t\bar{t}W}) = (8.5\%, 7.5\%)$ as reference values, obtained by halving the current theoretical uncertainties on the SM predictions~\cite{Frederix:2017wme}. The resulting $95\%$ CL bound with $L = 30 \ab^{-1}$ is
\begin{equation} \label{eq:SSL_bounds} 
\qquad\qquad\qquad \Lambda / \sqrt{|\ctt| } \,>\, 6.1 \TeV\;\,(\mathrm{no}\;\mathrm{syst.\hspace{-1mm}:}\;6.9 \TeV) \,. \qquad \quad \qquad (\mathrm{FCC}\mbox{-}\mathrm{hh},\,\mathrm{SSL})
\end{equation}
\begin{table}[t]
	\begin{center}
	\begin{tabular}{c|c|c|c}
		category&processes & decay channel & $\sigma \times \mathrm{BR}$ [fb]\\\hline
		\multirow{2}{*}{$t\bar{t}t\bar{t}$ (signal)} & $t\bar{t}t\bar{t}$ & \multirow{2}{*}{$W_{\ell^{+}}\:W_{\ell^{-}}\:W_{\ell^{\pm}}\: W_{\rm had}$} & \multirow{2}{*}{$0.206$}\\
		& $\Lambda / \sqrt{| c_{tt}| } = 6\;\mathrm{TeV}$ &  & \\\hline
		$t\bar{t}t\bar{t}$ (SM)&$t\bar{t}t\bar{t}$&$W_{\ell^{+}}\:W_{\ell^{-}}\:W_{\ell^{\pm}}\: W_{\rm had}$&$90.9$~\cite{Frederix:2017wme}\\\hline
		$t\bar{t}W$&$t\bar{t}W^{\pm}jj$ &$W_{\ell^{+}}\:W_{\ell^{-}}\:W_{\ell^{\pm}}$&63.4$^\dagger$\hspace{0.25mm}\cite{Torrielli:2014rqa}\\\hline
		\multirow{2}{*}{$t\bar{t}Z$} & $t\bar{t}Z\,+\,$0,1,2 jets &$W_{\ell^{\pm}}\:W_{\rm had}\:Z_{\ell^+ \ell^-}$&1120~\cite{Contino:2016spe}\\
		&$t\bar{t}Zjj$ &$W_{\ell^{+}}\:W_{\ell^{-}}\:Z_{\ell^+ \ell^-}$&82.6\\\hline
		\multirow{3}{*}{$t\bar{t}h$} &$t\bar{t}h,\: h\rightarrow WW^*$&$W_{\ell^{+}}\:W_{\ell^{-}}\:W_{ \ell^\pm}\:W_{\rm had}$&190~\cite{Contino:2016spe}\\
		&$t\bar{t}h,\: h\rightarrow ZZ^*$&$W_{\ell^{\pm}}\:W_{\rm had}\:Z_{\ell^+ \ell^-}\:Z_{\rm had}$&24.0~\cite{Contino:2016spe}\\
		&$t\bar{t}h,\: h\rightarrow \tau^+\tau^-$&$W_{\ell^{+}}\:W_{\rm \ell^-}\:\tau_{\ell^{\pm}}\:\tau_{\rm had}$&44.2~\cite{Contino:2016spe}\\\hline
		\multirow{2}{*}{other} &$tZbjj$&$W_{\ell^{\pm}}\:Z_{\ell^+ \ell^-}$&145\\
		&$t\bar{t}W^+W^-$&$W_{\ell^{+}}\:W_{\ell^{-}}\:W_{\ell^\pm}\:W_{\rm had}$&11.7\\\hline
		\multirow{2}{*}{fake $\ell$} &$t\bar{t}jj$&$W_{\ell^{+}}\:W_{\ell^{-}}$&$K_{t\bar{t}}\;4.63 \times 10^5$\\
		&$t\bar{t}bjj$&$W_{\ell^{+}}\:W_{\ell^{-}}$&$K_{t\bar{t}}\;1.06 \times 10^4$\\
	\end{tabular} 
	\end{center}
	\caption{3L signal and background processes at $\sqrt{s}=100 \TeV$. Samples with different jet multiplicities were merged using the MLM prescription with matching scale of $30 \GeV$. The cuts $p_T^j > 50 \GeV$ and $|\eta_j| < 5$ are imposed on jets arising from QCD radiation, but no cuts are applied to decay products of heavy particles. The subsequent baseline selection, discussed in Sec.~\ref{sec:3L}, requires $\geq 4$ jets among which $\geq 3$ are $b\hspace{0.2mm}$-tagged. The higher-order cross sections we use for normalization always assume $\mu = M_T/2$ (note that in~\cite{Frederix:2017wme} this is not the central choice for $t\bar{t}t\bar{t}$). The $^\dagger$ indicates that $p_T^j > 100 \GeV$ was exceptionally required, to match~\cite{Torrielli:2014rqa} (we have checked that this different initial cut has negligible impact on the event yield after the complete selection). Whenever they do not appear in $t\bar{t}$ or $b\bar{b}$ pairs, the symbols $t$ and $b$ refer to either particles or antiparticles. To the $t\bar{t}\,+\,$jets samples used to estimate the fake lepton background we apply a $K_{t\bar{t}} = 1.4$.}
	\label{tab:3L_bkg}
\end{table}
%

\subsection{Trileptons} \label{sec:3L}
In the trilepton channel the main backgrounds are the irreducible SM $t \bar{t} t \bar{t}$ production and the $t\bar{t}W$, $t\bar{t}Z$, $t\bar{t}h+\,$jets processes. The full list of backgrounds we consider is given in Table~\ref{tab:3L_bkg}, together with the MC generation-level cross sections. We generate processes giving rise to three leptons and at least four jets at the matrix element level. The fake lepton background is generated using the same method as in the SSL analysis of Sec.~\ref{sec:SSL}, but applied to a different set of processes. The $Q$flip background is negligible, since no requirement is imposed on the lepton charges. The event selection is analogous to the one for SSL: after lepton and jet candidates are identified as in Eq.~\eqref{eq:candidates}, we apply the same overlap removal procedure. In addition, the baseline selection requires
\begin{align}
&\qquad\quad\mathrm{exactly\; three\; leptons\;with\;} p_T^{\ell} > 25 \GeV \,, \nonumber \\ 
&\geq 4 \;\mathrm{jets}\,, \; \mathrm{of\;which}\; \geq 3\; b\hspace{0.2mm}\mbox{-}\mathrm{tagged}\,, \qquad H_T > 400 \GeV \,,
\end{align}
and events where among the three leptons appears one opposite-sign, same-flavor lepton pair satisfying $|m_{\ell_i^+ \ell_i^-} - m_Z| < 15\GeV$ are vetoed, to suppress backgrounds containing a leptonic $Z$ decay. The requirement of three leptons with $p_T^\ell > 25 \GeV$ should allow for a straightforward triggering on these events. Notice that these selection requirements are orthogonal to those of the SSL analysis, which will ease the combination of the results.

\begin{figure}[!t]
    \centering
             \raisebox{0\height}{\includegraphics[width=0.495\textwidth]{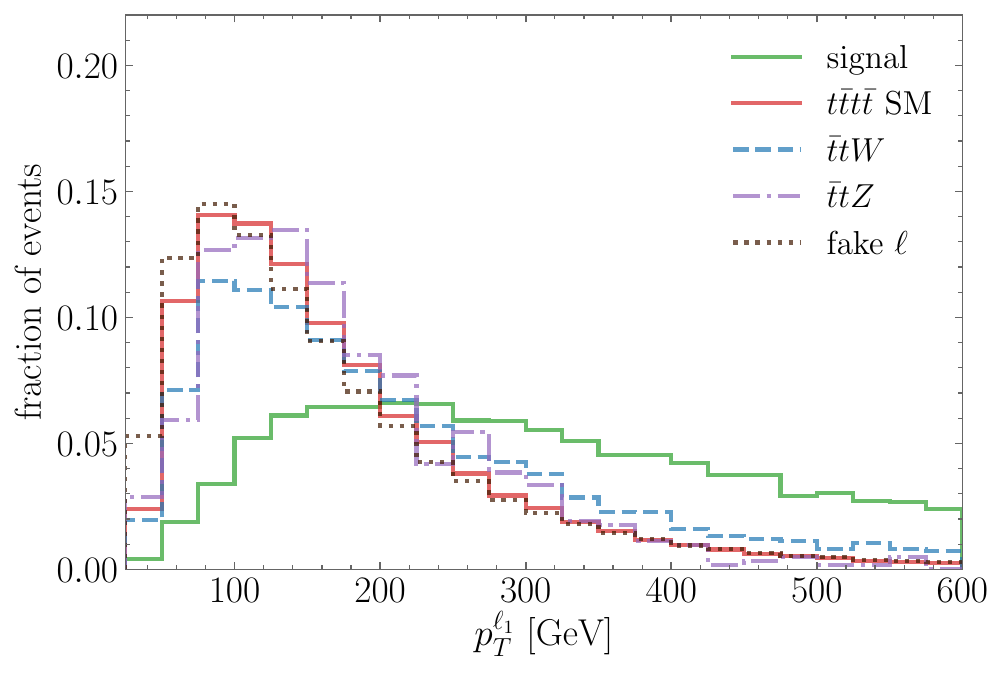}} 
     \raisebox{0.009\height}{\includegraphics[width=0.489\textwidth]{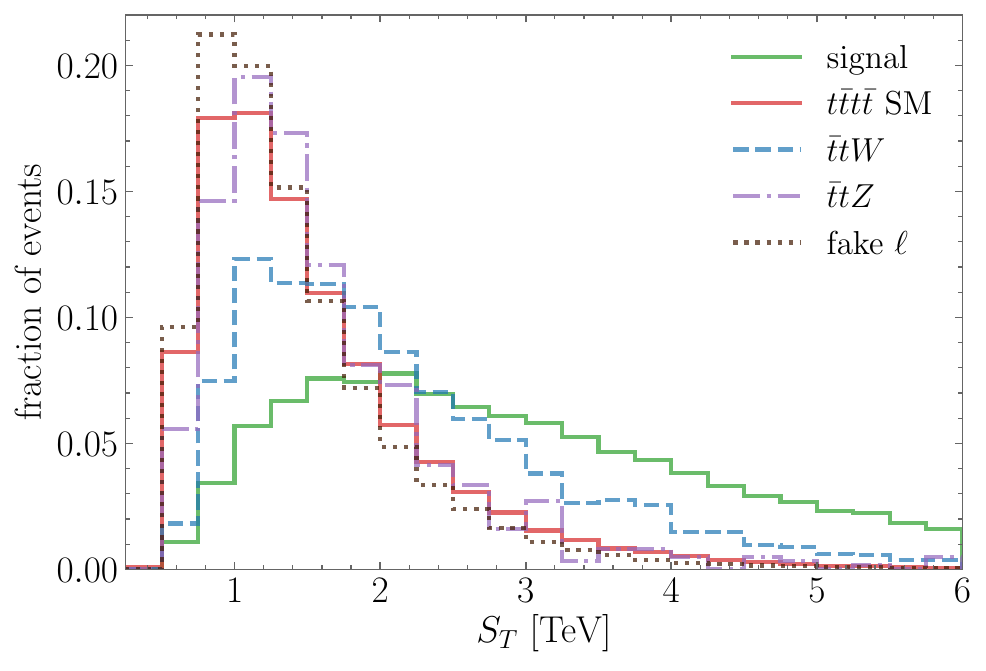}} 
    \caption{Normalized distributions of the leading lepton $p_T$ {(left)} and the scalar sum of the transverse momenta of all jets and the three leptons {(right)}, after the baseline 3L selection, for the signal and the main backgrounds.}
    \label{fig:3L_distrib}
\end{figure}
\renewcommand{\tabcolsep}{3.25pt}
\begin{table}[!t]
	\begin{center}
	\begin{tabular}{c|c|c|c|c|c|c|c|c|c}
		& $\mathrm{signal}$ & $t\bar{t}t\bar{t}$ & \multirow{2}{*}{$t\bar{t}W$} & \multirow{2}{*}{$t\bar{t}Z$}&\multirow{2}{*}{$t\bar{t}h$}&\multirow{2}{*}{other} & {fake} & $\mathcal{S}\;\mathrm{at}$ &  $S/B$ \\
		& $ \Lambda / \sqrt{|\ctt|} = 6 \TeV$ & {SM} &&&& & $\ell$ & $30/$ab & $[10^{-2}]$  \\\hline
		baseline&  21 & $ 6700$ & $570$ & $1400$ & $680$  & $250$ & $2100$ & 1.1 & 0.18 \\
		$p_T^{\ell_1 } > 275 \GeV$ & 13 & $1000$ & $160$ & $180$ & $74$ & $52$ & $310$ & 1.7 & 0.73 \\
		$S_T \in [3,4] \TeV$ & 2.9   & $160$  & 35  & 13  & 9.4  & 6.4 & 33  & 0.99  & 1.1   \\
		$S_T \in [4,5] \TeV$ & 2.1   & $66$  & 19 & 4.1  & 1.7  & 4.0  & 9.9  & 1.1  & 2.0   \\
		$S_T > 5 \TeV$ & 3.7  & $39$  & 16 & 4.1   & 1.4  & 1.9  & 6.9  & 2.4  & 5.3  \\
	\end{tabular} 
	\end{center}
	\caption{Cut flow for the 3L final state, with cross sections in ab.}
	\label{tab:3Lcut}
\end{table}

After the baseline selection, for a reference BSM scale $\Lambda / \sqrt{|\ctt| } = 6\TeV$, we have $S/B \sim 2 \times 10^{-3}$, as shown in Table~\ref{tab:3Lcut}. Normalized distributions of $p_T^{\ell_1}$ and $S_T$ at the baseline stage are shown in \Fig{fig:3L_distrib}. We adopt the same additional cuts as for the SSL selection, namely $p_T^{\ell_1} > 275 \GeV$ and $S_T > 3 \TeV$, and divide the remaining events into three $S_T$ bins, with $S/B$ in the $(1\,$-$\,5)\times 10^{-2}$ range. We thus obtain at $95\%$ CL
\begin{equation} \label{eq:3L_bounds} 
\qquad\qquad\qquad \Lambda / \sqrt{|\ctt|} \,>\, 5.8 \TeV\;\,(\mathrm{no}\;\mathrm{syst.\hspace{-1mm}:}\;6.6 \TeV)\,, \qquad \quad \qquad (\mathrm{FCC}\mbox{-}\mathrm{hh},\,\mathrm{3L})
\end{equation}
where $L = 30 \ab^{-1}$ was assumed.

\subsection{Same-sign dileptons and trileptons combination and discussion}
\begin{figure}[t]
    \centering
     \raisebox{0\height}{\includegraphics[width=0.49\textwidth]{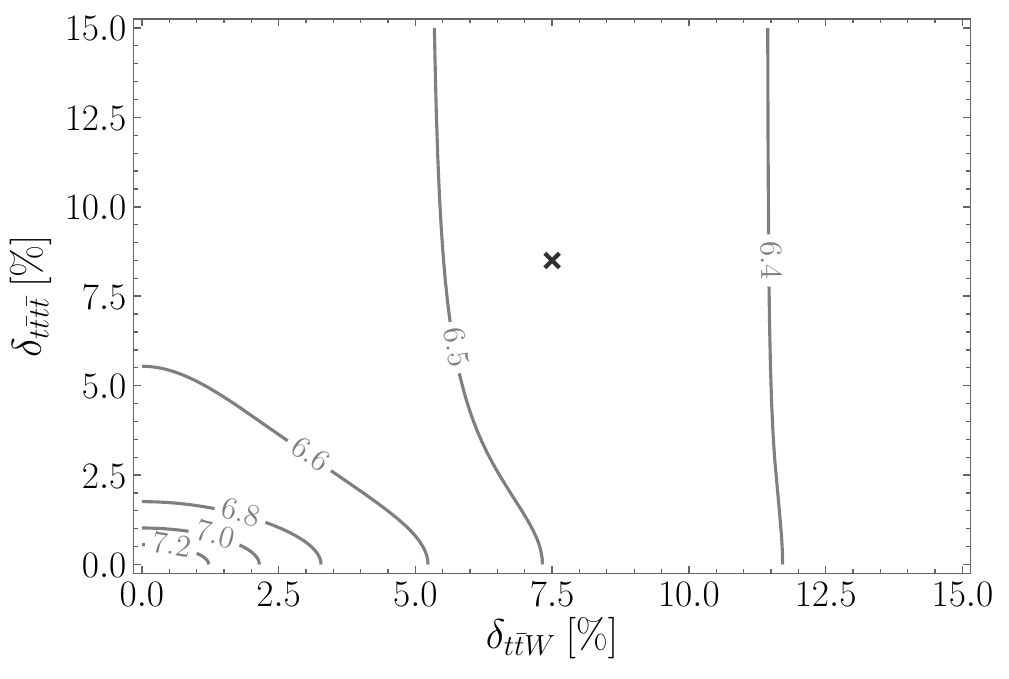}} \hspace{0.2mm}
     \raisebox{+0.003\height}{\includegraphics[width=0.49\textwidth]{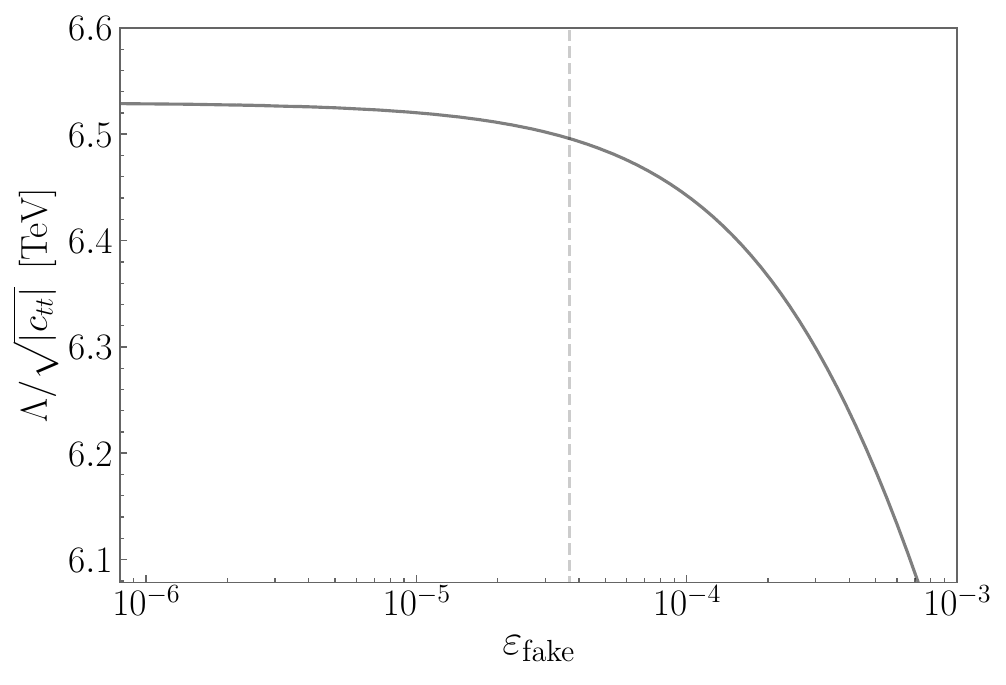}}
    \caption{{\it Left:} contours of the $95\%$ CL bound on $\Lambda / \sqrt{|\ctt|}$ in TeV, obtained by combining the SSL and 3L analyses, in the plane of systematic uncertainties on the two main SM backgrounds. The cross indicates our reference values. {\it Right:} Impact on the bound of varying the probability for a jet to give rise to a fake lepton. The dashed line indicates our baseline assumption.}
    \label{fig:FCCsystematics}
\end{figure}

We now combine the results in the SSL and 3L final states, by considering a joint $\chi^2$ with $6$ orthogonal bins. We obtain
\begin{equation} \label{eq:joint_bounds}
\quad\quad\qquad \Lambda / \sqrt{|\ctt| } \,>\, 6.5 \TeV\;\,(\mathrm{no}\;\mathrm{syst.\hspace{-1mm}:}\;7.3 \TeV) \,, \qquad \quad \quad (\mathrm{FCC}\mbox{-}\mathrm{hh},\,\mathrm{SSL+3L})
\end{equation}
from $L = 30 \ab^{-1}$ and with the reference systematic uncertainties $(\delta_{t\bar{t}t\bar{t}}, \delta_{t\bar{t}W}) = (8.5\%, 7.5\%)$. The impact of varying these uncertainties is shown in the left panel of \Fig{fig:FCCsystematics}; we stress that we assume full correlation of each uncertainty across bins. In the right panel of \Fig{fig:FCCsystematics} we display the dependence of the combined bound on the fake lepton probability, whose value at FCC-hh is unknown and which we have fixed based on a fit to LHC data. The $Q$flip background affects only the SSL analysis and is $3\,$-$\,4$ times smaller than the fake lepton background in our benchmark scenario, so its impact remains small for any reasonable choice of the electron charge-flip probability $\varepsilon_{\rm flip}$.
\begin{figure}[!t]
    \centering
    \includegraphics[width=0.5\textwidth]{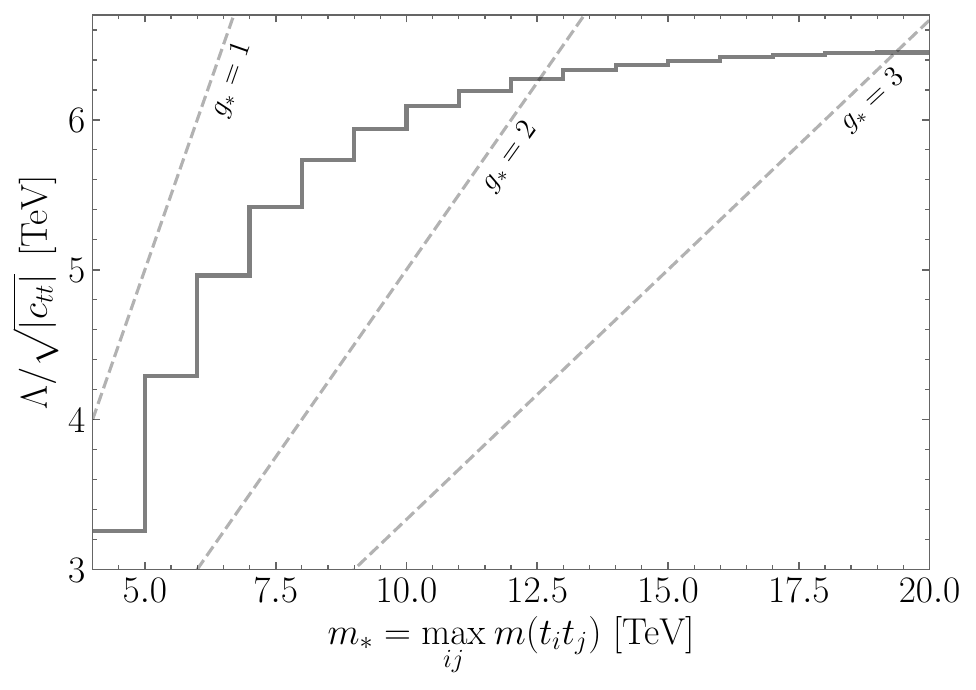}
    \caption{Combined SSL$\,$+$\,$3L bound as a function of $m_\ast$, the maximal allowed parton-level invariant mass of any pair of top or antitop quarks. Dashed lines indicate contours of constant new-physics coupling $g_\ast = \sqrt{|\ctt|}\, m_\ast / \Lambda$.}
    \label{fig:EFTvalidity}
\end{figure}

In addition, we want to ensure that our bounds arise from regions of phase space where the EFT expansion is under control. For this purpose we show in \Fig{fig:EFTvalidity} the combined SSL$\,$+$\,$3L bound on $\Lambda/\sqrt{|\ctt|}$ obtained by discarding events where the largest parton-level invariant mass of a top quark pair is larger than $m_\ast$, which represents the mass of new resonances. Since it is not possible to tell on an event-by-event basis whether the hard scattering involved a $t\bar{t}$, $tt$, or $\bar{t}\bar{t}$ pair, we make the conservative choice to discard events where the largest invariant mass of {\it any} such combination is larger than $m_\ast$. 

We now return to the role of the interference between the signal and SM $t\bar{t}t\bar{t}$ amplitudes. To quantify it, it is enough to work at the parton level, hence as rough proxy for our signal region we consider the process $pp\to t\bar{t} t\bar{t}$, followed by SSL decays and including the cuts $p_{T}^{\ell_1} > 200 \GeV$ and $H_T > 2 \TeV$.\footnote{For this check we only consider the dominant $O(\alpha_s^2)$ component of the SM amplitude.} We find the leading order cross section
\begin{equation}
\sigma(t\bar{t} t\bar{t})\,[\mathrm{fb}] = 1.5 + (0.3 \pm 0.3) \times 10^{-3} \, \frac{(6 \TeV)^2}{\Lambda^2/\ctt}  + 0.071 \, \frac{(6 \TeV)^4}{\Lambda^4/\ctt^2} \,,
\end{equation} 
where the coefficients are obtained by fitting to a set of cross sections calculated for varying $\ctt/\Lambda^2$, and the uncertainties on the SM and $O(\ctt^2)$ terms are negligible compared to the one on the linear term. This result confirms that interference can be safely neglected. The same conclusion applies to the 3L final state.
\begin{figure}[t]
    \centering
    \includegraphics[width=0.65\textwidth]{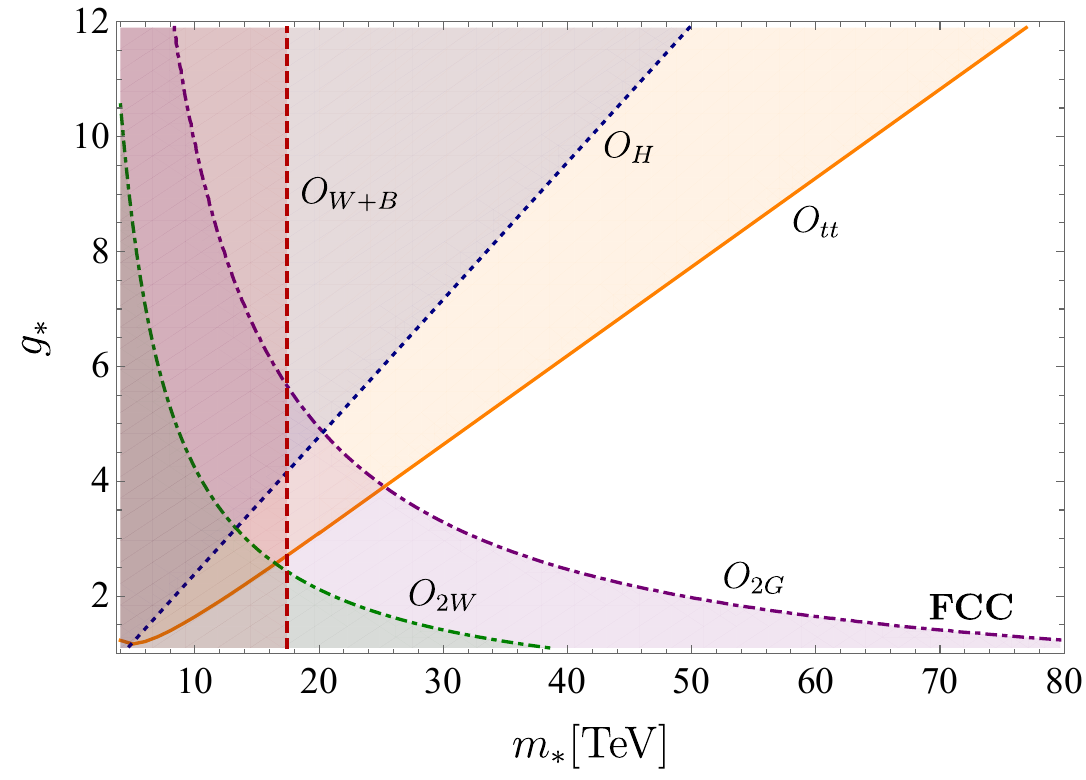}
    \caption{Future sensitivity at the FCC, including the ee/eh/hh stages, at 95\% CL in the $(m_*,g_*)$ plane of scenarios featuring a strongly-interacting Higgs and (right-handed) top quark. The different limits are associated with constraints on individual operators, each dominating the corresponding observables in a certain region of parameter space (see main text for details). The limit on $O_{tt}$ is derived using only FCC-hh.}
    \label{fig:FCCbounds}
\end{figure}

Finally, in Fig.~\ref{fig:FCCbounds} we show the impact of our combined SSL$\,+\,$3L bound, \Eq{eq:joint_bounds}, on the $(m_\ast, g_\ast)$ parameter space of CH models, and compare it with other, complementary probes which will become available throughout the development of the FCC program. 

Strikingly, four-top production at the FCC-hh provides the dominant sensitivity on the compositeness scale, $f|_{tt}^\fcc \gtrsim 6.5 \TeV$, outperforming tests of Higgs coupling deformations associated with $O_H$, as combined in~\cite{deBlas:2019rxi} which includes input from the HL-LHC and the FCC-ee, -eh, and -hh phases, resulting in $f|_{H}^\fcc \gtrsim 4.2 \TeV$  at 95\% CL. 
In addition, we show the projected constraint on $O_{W}$, $O_{B}$~\cite{deBlas:2019rxi}, namely $m_\ast > 17 \TeV$ at 95\% CL, as well as the expected FCC-hh bounds on $O_{2W}$ and $O_{2G}$, derived from charged- and neutral-current dilepton production~\cite{deBlas:2019rxi} (see also~\cite{Farina:2016rws}), and high-energy dijet and inclusive jet production~\cite{Alioli:2017jdo}, respectively. These observables dominate the sensitivity for moderate strength of the new-physics coupling $g_\ast$. Finally, we mention that strong constraints are also expected from CP-violating observables: the limit on $\widetilde O_\gamma$ from the future measurement of the electron EDM by the ACME III experiment~\cite{Doyle:2016} reaches $m_\ast > 115 \TeV$ at $95\%$ CL. However, this probe is left out of \Fig{fig:FCCbounds} due to its inherently different nature, as already done in \Fig{fig:LHCbounds}.

\subsection{Fully hadronic final state}
Finally, we turn to the signature that arises when all four tops decay hadronically. This channel benefits from a large branching ratio of $20\%$ and is intrinsically interesting because at the FCC-hh the hadronic tops will frequently possess multi-TeV transverse momenta, entering a kinematic regime that is only marginally accessible at the LHC (for which the fully hadronic signature was discussed in~\cite{Kim:2016plm}, albeit assuming a resonant signal). While this happens already in the SM, the relative importance of ultra-boosted tops increases further in the presence of heavy new physics that generates $O_{tt}$. To obtain a first estimate of the reach, we perform a crudely simplified analysis that requires four top-tagged jets, relying on the performance of existing hadronic top tagging methods developed for the LHC, as studied by CMS~\cite{CMS:2016tvk}. As a first step, we generate the signal and the main backgrounds, which are $t\bar{t}t\bar{t}$, $t\bar{t}jj$ and $jjjj$ production in the SM, at parton level with a $p_T > 200$~GeV cut on each undecayed top or jet. The interference between the BSM and SM four-top amplitudes is neglected, since we are interested in the high-energy regime. We then include the branching ratios for hadronic top decays and apply, on an event-by-event basis, the $p_T$-dependent efficiencies and mistag rates extracted from~\cite{CMS:2016tvk}.\footnote{For $200 \GeV < p_T < 600 \GeV$ ($p_T > 600 \GeV$) we use the low-$p_T$ (high-$p_T$) working point in Fig.~10 (Fig.~11) of~\cite{CMS:2016tvk}, assuming the combination of jet substructure algorithms corresponding to the light green points. The efficiency and mistag rate are assumed constant for $p_T > 1.5 \TeV$.} Finally, we select highly energetic events by requiring the total invariant mass of the four final-state objects to be larger than $5.5 \TeV$ and the sum of the transverse momenta to be larger than $4.5 \TeV$. Demanding $S/\sqrt{B} > 1.96$ for $L = 30 \ab^{-1}$ we find the $95\%$ CL bound 
\begin{equation} \label{eq:bound_allhad}
\quad\quad\quad\qquad\qquad \Lambda/\sqrt{|\ctt| } > 6.0 \TeV \,.  \qquad \quad \quad (\mathrm{FCC}\mbox{-}\mathrm{hh},\,\mathrm{fully\;hadronic, \,estimate})
\end{equation}
The corresponding signal cross section is $\approx 1.0 \ab$ and $S/B \approx 0.13$, which justifies omitting systematic uncertainties in first approximation. The background is dominated by SM four-top production with an $O(10)\%$ contribution from $t\bar{t}jj$, while $jjjj$ is negligible.

The estimate \Eq{eq:bound_allhad}, although obtained by means of rough approximations, indicates a promising potential for the fully hadronic channel. However, requiring a large $p_T$ for all four tops, as necessary in order to apply the results of~\cite{CMS:2016tvk}, severely suppresses the signal rate, ultimately limiting the sensitivity. This motivates pursuing a different strategy, where the two hardest tops are tagged using jet substructure algorithms whereas the two softest tops are identified from their resolved decay-product jets; this is in consonance with the topology of our signal, which is characterized by a high-energy $t\bar{t} \to t\bar{t}$ scattering mediated by $O_{tt}$. The challenge of this approach is to retain a strong rejection capability against the $t\bar{t}jj$ background, in particular the configuration where the two tops have larger $p_T$'s than the light jets', in which case the signal/background discriminating power must be obtained from the ``soft'' component of the event. 

To study this problem, we generate SM four-top production and $t\bar{t}jj$ with hadronic top decays, using the MadGraph5$\char`_\hspace{0.15mm}$aMC@NLO$\,$--$\,$Pythia8$\,$--$\,$Delphes3 chain. All final-state partons are required to have $p_T > 100 \GeV$, whereas the leading (subleading) jet is required to have $p_T > 900\,(800) \GeV$ and the $H_T$ must exceed $2 \TeV$. The only notable setup differences compared to the multilepton analyses are that we use the default Delphes card and set $R = 0.3$ for the (anti-$k_t$) jet clustering, because using such narrower jets allows for a more efficient matching of the hadron- and parton-level objects, therefore easing the isolation of a $t\bar{t}jj$ sub-sample containing the configuration where the light jets are softer than the tops (which happens for $O(10)\%$ of the events). Two different strategies are investigated to separate this background from SM $t\bar{t}t\bar{t}$ production: one based on top invariant mass reconstruction, and one employing a neural network discriminant. 

For the first strategy we implement an algorithm which first removes the two hard jets that are matched to partonic tops, and then identifies two sets of up to three jets each, whose invariant masses are closest to $m_t$, with each set required to contain at least one $b$-tagged jet.  This method results in an $8\%$ efficiency per event on SM four-top production and $0.4\%$ on the background. For the second strategy we use the same MC samples to train a three-layer neural network with $2910$ neurons per layer, which takes as input features the $p_T$, $\eta$, $\phi$, mass, number of tracks, and $b$-tag flag of up to $26$ jets with $p_T > 50 \GeV$ in each event (including, in particular, the two hard jets which are matched to tops), as well as information on possible additional particles such as taus and photons, and on missing transverse energy. At the optimal threshold value the efficiency on SM four-top production is $16\%$, significantly higher than for the mass reconstruction procedure, but this comes at the price of a less effective background suppression of $4\%$. The above efficiencies are obtained neglecting systematic uncertainties. 

Unfortunately, neither approach yields a satisfactory combination of signal efficiency and background rejection, resulting in weaker bounds on $\Lambda/\sqrt{| \ctt | }$ than the estimate in \Eq{eq:bound_allhad}. 
Nevertheless, we believe that our attempts have only scratched the surface of the fully hadronic four-top final state, while uncovering some of the main obstacles that need to be overcome. The sensitivity of this channel is thus still waiting to be untapped, for instance through the development and application of FCC-tailored and/or machine learning-based top tagging algorithms (see e.g.~\cite{Jamin:2019mqx,Kasieczka:2019dbj,Marzani:2019hun}) that encompass both the resolved and boosted top regimes.
Judging from our preliminary estimates, this channel has the potential to give the strongest constraint on the new-physics scale at the FCC-hh, further improving on our multilepton results.


\section{Future electron-positron colliders} \label{sec:lepton}

\begin{figure}[!t]
    \centering
    \includegraphics[width=0.55\textwidth]{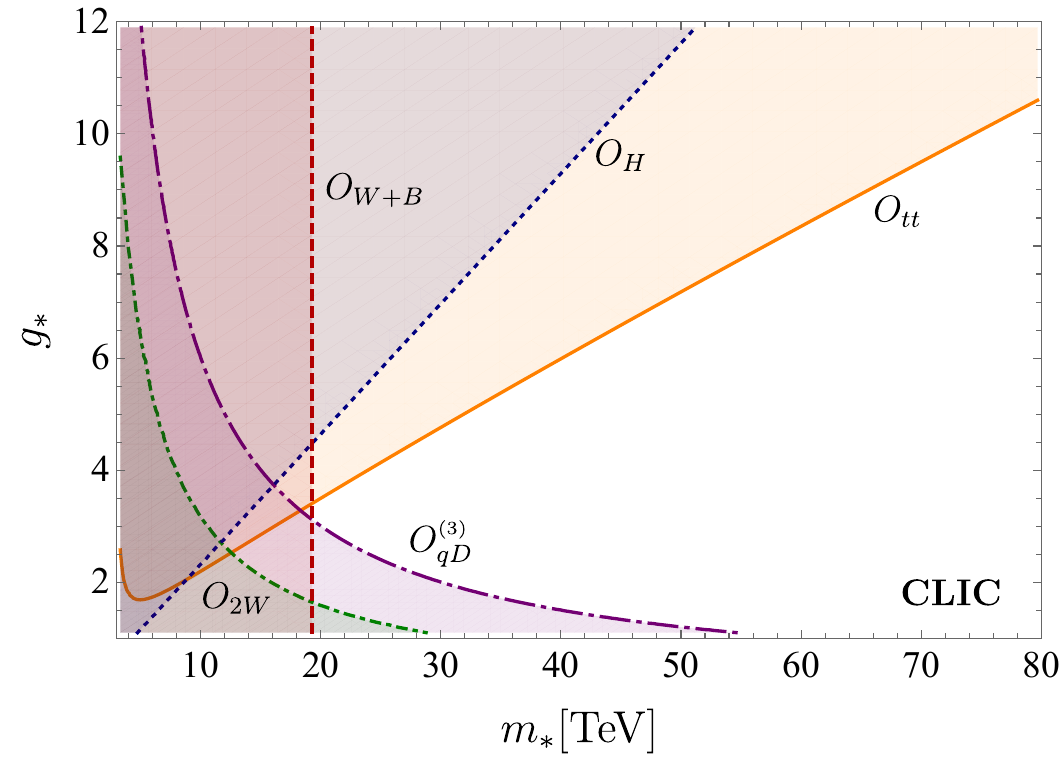}
    \vspace{0.15cm}
    \includegraphics[width=0.55\textwidth]{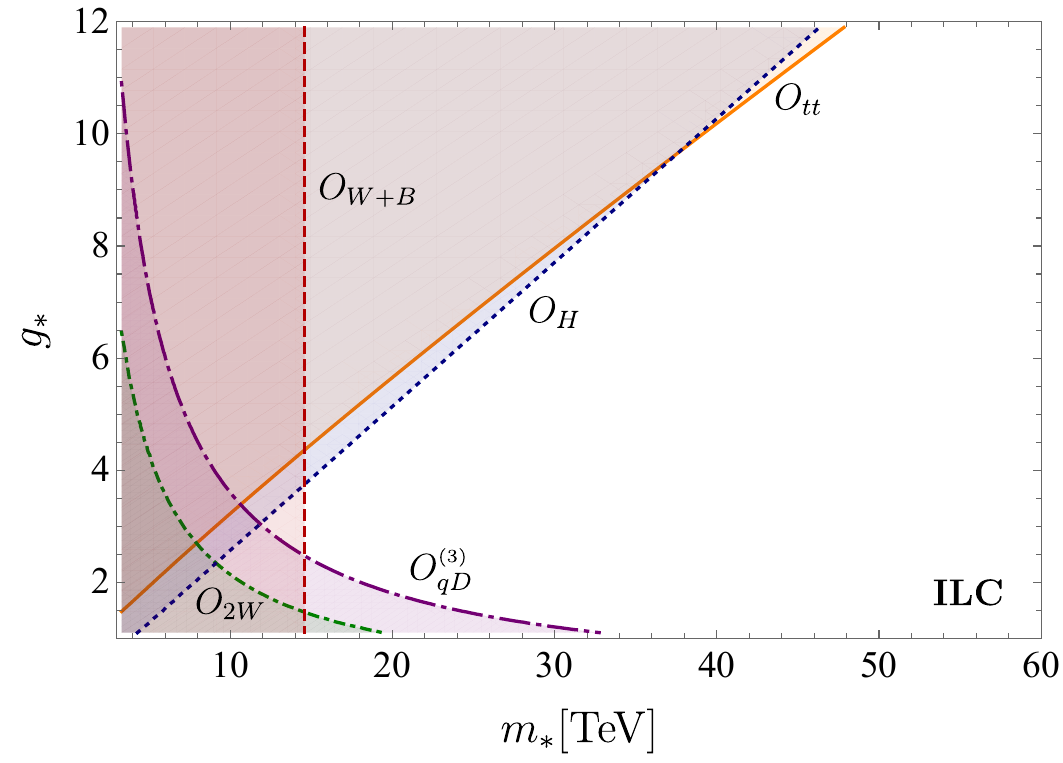}
    \vspace{0.15cm}
    \includegraphics[width=0.55\textwidth]{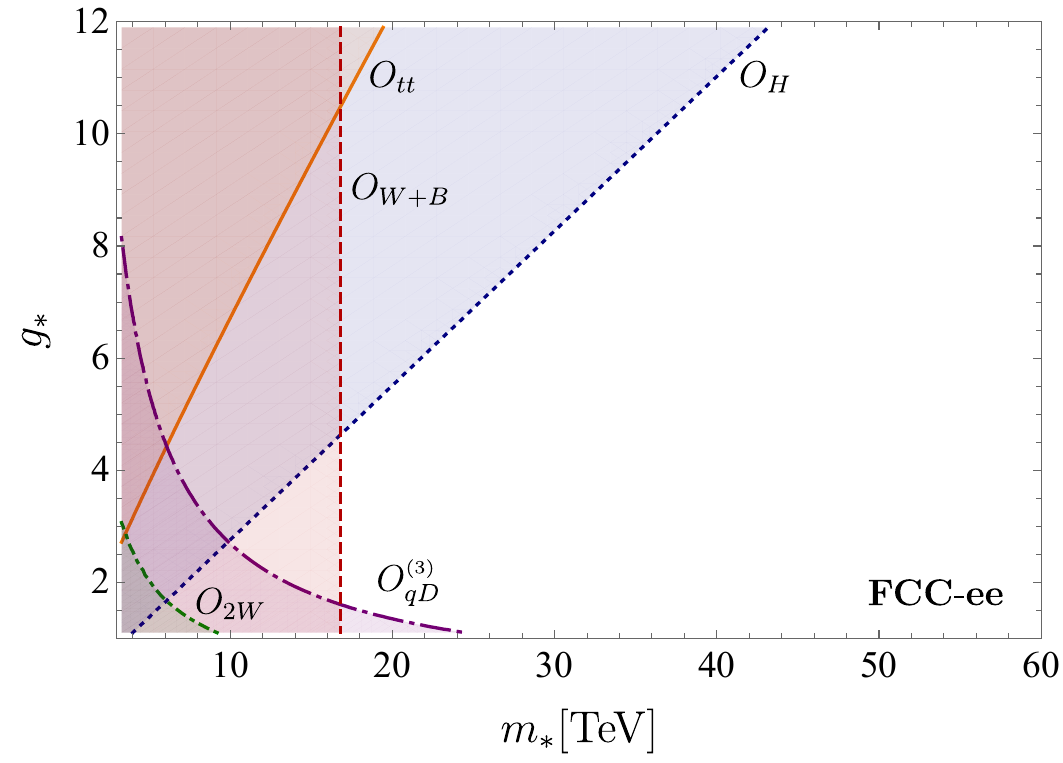}
    \caption{Future sensitivities at lepton colliders: CLIC (top), ILC (middle), and FCC-ee (bottom), at 95\% CL in the $(m_*,g_*)$ plane of scenarios featuring a strongly-interacting Higgs and (right-handed) top quark. The different limits are associated with constraints on individual operators, each dominating the corresponding observables in a certain region of parameter space (see main text for details).}
    \label{fig:LCbounds}
\end{figure}

In this section we show that future leptonic machines have much to inform on the fate of a strongly-interacting top quark. The colliders under consideration are CLIC \cite{Charles:2018vfv}, the International Linear Collider (ILC) \cite{Bambade:2019fyw}, and the FCC-ee \cite{Abada:2019lih}.
We will not be carrying out any new analysis towards the extraction of their sensitivity to the dimension-six effective operators of interest, since this has been the subject of a number of detailed and comprehensive studies. Instead, we merely yet crucially reinterpret the relevant results in terms of the expected effects associated with a strongly-interacting (right-handed) top quark, in particular via the four-top operator \Eq{o4tR}.\footnote{To some extent, our analysis resembles that of \cite{Durieux:2018ekg}. However, as in previous sections, we focus on a single operator at a time, the one leading to the largest sensitivity in a given region of the $(m_*,g_*)$ parameter space, which is not always the same operator as claimed in that study. Besides, by considering exclusive constraints, we avoid issues associated with cancellations from different operators in a given observable.}
The different collider specifications can be found in the pertinent works:~\cite{Durieux:2018tev} for what regards the top sector, and~\cite{deBlas:2019rxi} concerning universal effects, which we use to draw a comparison of both types of probes in the context of composite Higgs models. The runs from where most of the sensitivity to a composite top comes from are those at the highest energies:~$\sqrt{s} = 3 \TeV$ ($L = 3 \ab^{-1}$) at CLIC, $\sqrt{s} = 1 \TeV$ ($L = 1 \ab^{-1}$) at ILC, and $\sqrt{s} = 365 \GeV$ ($L = 1.5 \ab^{-1}$) at FCC-ee.\footnote{Notice the mildly different assumptions made for the luminosities and energies of these machines in~\cite{Durieux:2018tev} and~\cite{deBlas:2019rxi}.}
The reason for this is that at linear colliders the best process to probe such type of physics is top-pair production, $e^+ e^- \to t \bar t$. In our new-physics oriented analysis we find that the largest effects are associated with the four-fermion operators
\beq
\frac{\cte}{\Lambda^2} (\bar e_R \gamma_\mu e_R) (\bar t_R \gamma^\mu t_R) + \frac{\ctl}{\Lambda^2} (\bar \ell_L \gamma_\mu \ell_L) (\bar t_R \gamma^\mu t_R)\,,
\label{eett}
\eeq
where both $e_R$ and $\ell_L$ correspond to first-generation leptons.
Since we consider a negligible degree of lepton compositeness, as motivated by their small Yukawa couplings, the largest contribution to the coefficients in \Eq{eett} arises from operators of the form of \Eq{oDtR}, in particular from $O_{tD}$ which, given the equation of motion \Eq{Beom}, yields $\cte = g' \ctD$ and $\ctl = g'\ctD/2$. 
What is important to notice is that at the relevant scale, $\mu = \sqrt{s}$, the coefficient of $O_{tD}$ is dominated by the RGE contribution from the four-top operator $O_{tt}$,
\beq
\ctD(\mu)= \ctD(m_*) + \ctt(m_*) \frac{32}{9} \frac{g'}{16\pi^2} \log \Big(\frac{m_*^2}{\mu^2}\Big) \,,
\label{rgectD}
\eeq
for a mildly strong coupling $g_*$, since recall $\ctD/\Lambda^2 \sim g'/m_*^2$ and $\ctt/\Lambda^2 \sim g_*^2/m_*^2$ at the scale $m_*$, where the coefficients are generated. Therefore, a strongly-interacting (right-handed) top quark leads to a new-physics amplitude that scales like
\beq
\mathcal{M}_{e^+ e^- \to t \bar t} \sim \frac{g'^2}{16\pi^2} \frac{s}{f^2} \log \Big(\frac{m_*^2}{s}\Big) \,.
\label{Aeettt}
\eeq
From the expected $1\sigma$ sensitivity to the operator $O_{te}$ at the 3 TeV CLIC, $\cte/\Lambda^2 < 1.6 \times 10^{-4} \TeV^{-2}$ \cite{Durieux:2018tev},\footnote{The experimental sensitivity to $O_{t\ell}$ is similar, but we neglect it in setting the limit because $\ctl = \cte/2$.} 
we arrive at the 95\% CL bound on the compositeness scale
\beq
f|^{\clic}_{tt} > 7.7 \TeV \, ,
\label{fboundCLIC}
\eeq
for $m_* = 4 \pi f$ (to fix the size of the logarithm in \Eq{Aeettt}). This is stronger than the expected sensitivity to be achieved in Higgs measurements via the operator $O_H$, $f|^{\clic}_H > 4.3 \TeV$ \cite{deBlas:2019rxi}, also shown in \Fig{fig:LCbounds} in the $(m_*,g_*)$ plane.
At the ILC the sensitivity via the four-top operator is comparatively lower, $f|^{\ilc}_{tt} > 4.1 \TeV$ ($\cte/\Lambda^2 < 7 \times 10^{-4} \TeV^{-2}$ \cite{Durieux:2018tev}), yet similar to that from the Higgs. 
Finally, the importance of high collision energies for this type of probes is reflected in FCC-ee bounds on $\cte$, $\ctl$, which are approximately an order of magnitude weaker, yielding a significantly lower sensitivity $f|_{tt}^\fccee > 1.6 \TeV$ ($c_{te} < 4.3 \times 10^{-3} \TeV^{-2}$~\cite{Durieux:2018tev}), see \Fig{fig:LCbounds}.

Let us note at this point that our analysis of one operator at a time must be interpreted with a certain care, particularly in the case where several operators enter a given process. For instance, while the one-loop contribution from $O_{tt}$ gives the leading non-standard effect in $e^+ e^- \to t \bar t$ at large $g_*$, for small new-physics couplings other operators become comparable and eventually dominate, in particular the finite contribution to $O_{tD}$ generated at $m_*$, see Eq.~\eqref{rgectD} (loops from other four-top operators in \Eq{o4qL} are not enhanced by the strong coupling and thus always subleading). 
This implies that in the transition region cancellations could take place, reducing the sensitivity to new physics. Fortunately, this is not an issue that prevents us from probing those regions of parameter space, since they are tested in other processes via independent operators; specifically, tests of the operator $O_{W+B}$ in electroweak precision data are expected to provide at CLIC the bound $m_* > 19 \TeV$ at 95\% CL \cite{deBlas:2019rxi}, independent of the new-physics coupling. The same holds at ILC and FCC-ee, even though, as shown in \Fig{fig:LCbounds}, the sensitivity to the resonance scale is somewhat lower. 

The power of tests of the top sector in covering the parameter space of CH models goes beyond top-pair production. As already noted in~\cite{Durieux:2018ekg}, production of left-handed bottom pairs at lepton colliders is sensitive to effects that are enhanced at weak coupling, for instance via the operator $O_{qD}^{\text{\tiny (3)}}$ in \Eq{oDqL} with $\cqDt/\Lambda^2 \sim (y_t/g_*)^2 g/m_*^2$, which from the equations of motion contributes as a contact term to the amplitude for $e^+ e^- \to b \bar b$. As we show in \Fig{fig:LCbounds}, this is superior to electroweak precision tests in the form of the $\Wp$ parameter, to be measured in e.g.~$e^+ e^- \to \mu^+ \mu^-$, because the coefficient of $O_{qD}^{\text{\tiny (3)}}\,$ is enhanced by the top Yukawa coupling, $\cqDt/( g\hspace{0.3mm} c_{2W}) \sim (y_t/g)^2$, while the experimental precision in the two processes is expected to be comparable. In addition, it is worth noting that at CLIC and ILC, bottom-pair production could provide a non-negligible sensitivity to the masses of the composite resonances, independently of $g_*$, because of RGE effects associated with the four-top operator $O_{tq}$ in \Eq{o4qL},
\beq
\cqD(\mu)= \cqD(m_*) + \ctq(m_*) \frac{g'}{12 \pi^2} \log \Big(\frac{m_*^2}{\mu^2}\Big) \,.
\label{rgecqD}
\eeq
Given that $\ctq/\Lambda^2 \sim y_t^2/m_*^2$, we find $m_*|_{tq}^{\clic} > 6.5 \TeV$, a significant constraint, yet weaker than the sensitivity to be achieved from the $\Sp$ parameter ($O_{W+B}$).

Let us finally comment on the potential sensitivity from measurements of anomalous top and bottom couplings to the $Z$ boson. Under our assumptions, both the corrections to the $Zt_Rt_R$ and $Zb_Lb_L$ couplings, dominated by $O_{Ht}$ and $O_{Hq} + O_{Hq}^{\text{\tiny (3)}}$ respectively, do not receive large tree-level contributions, being protected by a $P_{LR}$ symmetry. This then implies that the dominant contributions arise from the RGE associated with $O_{Hq}$ and $O_{Hq}^{\text{\tiny (3)}}$ themselves and with the leading four-top operators in $g_*$, $O_{tt}$ and $O_{tq}$,
\begin{align}
c_{Ht}(\mu) & \simeq \ctt(m_*) \frac{y_t^2}{2 \pi^2} \log \Big(\frac{m_*^2}{\mu^2}\Big) \,,
\label{rgecHt} \\
c_{Hq}(\mu) + c_{Hq}^{\text{\tiny (3)}}(\mu) & \simeq [3 \ctq(m_*) + 4 c_{Hq}^{\text{\tiny (3)}}(m_*) ] \frac{y_t^2}{16 \pi^2} \log \Big(\frac{m_*^2}{\mu^2}\Big) \, ,
\label{rgecHq}
\end{align}
where we set $c_{Hq}(m_\ast) + c_{Hq}^{\text{\tiny (3)}} (m_\ast) \simeq 0$ and neglected gauge coupling terms, which are relatively suppressed by $(g/y_t)^2$ \cite{Elias-Miro:2013mua}.
We find that the expected precision on these couplings \cite{Durieux:2018tev,deBlas:2019rxi} is not high enough to give rise to any constraint at the level of those already discussed. In fact, not even measurements of the $Zt_Lt_L$ coupling, which receives relatively large corrections $(c_{Hq}-c_{Hq}^{\text{\tiny (3)}})/\Lambda^2 \sim y_t^2/m_*^2$ and for which the prospective exclusive $1\sigma$ bound is e.g.~at the ILC $0.075 \TeV^{-2}$ \cite{Durieux:2018tev}, can compete with universal probes.
Dropping $P_{LR}$ symmetry, i.e.~for $c_{Ht}/\Lambda^2 \sim 1/f^2$ and $(c_{Hq}+c_{Hq}^{\text{\tiny (3)}})/\Lambda^2 \sim y_t^2/m_*^2$, the situation at CLIC and ILC is actually not much different. 
For instance, at the ILC $c_{Ht}/\Lambda^2 < 0.15 \TeV^{-2}$ at 95\% CL \cite{Durieux:2018tev}, which leads to $f|_{Ht}^{\ilc} > 2.6 \TeV$, a weaker sensitivity than from $e^+ e^- \to t \bar t$, even though the latter is loop suppressed. Likewise, from $(c_{Hq}+c_{Hq}^{\text{\tiny (3)}})/\Lambda^2 < 0.019 \TeV^{-2}$ \cite{deBlas:2019rxi} we find $m_*|_{Hq}^{\clic} > 10 \TeV$, lower than the scale to be reached with electroweak precision data.
At the FCC-ee instead the absence of $P_{LR}$ would make a difference since, as we discussed, the lower $\sqrt{s}$ penalizes the effects of contact interactions. We find $f|_{Ht}^{\fccee} > 1.8 \TeV$, which however is still below the expected compositeness scale probed with Higgs measurements, while $m_*|_{Hq}^{\fccee} > 24 \TeV$, under optimistic assumptions on the systematics of the bottom forward-backward asymmetry \cite{deBlas:2019rxi}.


\section{Conclusions} \label{sec:conclusions}

In this paper we have shown that some of the proposed high-energy colliders have an outstanding sensitivity to four-top operators, which constitute telltale signs of a strongly-interacting top quark. Focusing on the $O_{tt} = (\bar{t}_R \gamma_\mu t_R)^2$ operator (with coefficient $\ctt/\Lambda^2$), we have performed realistic analyses of four-top production at the FCC-hh in the same-sign dilepton and trilepton final states, and inspected the fully-hadronic final state. We have also reinterpreted previous results to constrain $O_{tt}$ at future high-energy lepton colliders, through its one-loop renormalization group contributions to top-pair production. We have obtained the following $95\%$ CL bounds,
\begin{align}
& \!\!\!\!\!\!\!\!\!\!\!\! \textrm{FCC-hh}_{\, 100 \TeV, \,\, 30 \ab^{-1}}^{\, pp \,\to\, t\bar{t}t\bar{t}}: \qquad \quad \Lambda/{\sqrt{|\ctt|}} > 6.5 \TeV \,, \nonumber \\
& \!\!\!\!\!\!\!\!\!\!\!\! \textrm{CLIC}_{\, 3 \TeV, \,\, 3 \ab^{-1}}^{\, e^+ e^- \to\, t\bar{t}}: \qquad \qquad \quad \: \Lambda/{\sqrt{|\ctt|}} > 7.7 \TeV \,, \\
& \!\!\!\!\!\!\!\!\!\!\!\! \textrm{ILC}_{\, 1 \TeV, \,\, 1 \ab^{-1}}^{\, e^+ e^- \to\, t\bar{t}}: \qquad \qquad \quad \;\;\:\, \Lambda/{\sqrt{|\ctt|}} > 4.1 \TeV \,. \nonumber
\end{align}
For context, the $13\TeV$ LHC limit as derived from a combination of $t\bar{t}t\bar{t}$ final states is $\Lambda/{\sqrt{|\ctt|}} > 0.73 \TeV$, based on approximately $36 \fb^{-1}$. Thus, a tantalizing result of our study is that both the FCC-hh and CLIC at its highest-energy run would increase by an order of magnitude the reach on the scale of new physics. In contrast, the lower energy FCC-ee ($365 \GeV$, $1.5 \ab^{-1}$) displays a significantly milder reach of $\Lambda/{\sqrt{|\ctt|}} > 1.6 \TeV$.

In addition, we have studied the moderate excesses of events observed by ATLAS and CMS in their LHC Run 2 measurements of $t\bar{t}t\bar{t}$, $t\bar{t}W$, $t\bar{t}Z$, and $t\bar{t}h$ in multilepton plus jets final states. We have attempted a first interpretation of these results in the context of heavy physics beyond the SM, examining the latest CMS four-top search in terms of the operators $O_{tt}$ and $O_{Ht}$, the latter of which modifies the $Zt_R t_R$ coupling with respect to the SM. While far from conclusive, our analysis shows that a new-physics scale of around $0.75 \TeV$ could improve the agreement with multilepton$\,$+$\,$jets data, while remaining consistent with other competing measurements, notably in the Higgs sector. Further studies are warranted, both at the phenomenological level, including a more comprehensive set of measurements, and at the experimental level, where a complete modeling of the impact of higher-dimensional operators can be achieved. A well-motivated set would include, beyond $O_{tt}$ and $O_{Ht}$, the operator $O_{y_t}$, which controls non-standard contributions to the $htt$ coupling; these three operators are weakly constrained by other measurements yet their coefficients are expected to be large.\footnote{A more comprehensive set would incorporate $O_H$ and $O_{Hq}^{\text{\tiny (3)}}$ (with $c_{Hq} = - c_{Hq}^{\text{\tiny (3)}}$) as well, which modify respectively the $hVV$ and $Wt_Lb_L$ couplings. The former is of relevance in e.g.~$th$ production.} In general, the current status of top data provides additional motivation to investigate the new-physics scenarios discussed in this work. 

Looking ahead, our FCC-hh analysis of $O_{tt}$ in multilepton final states can be repurposed to derive the reach on other four-top operators, which may play a central role under different theoretical assumptions. On the other hand, exploiting the whole potential of the fully-hadronic signature requires a targeted study. Furthermore, the indirect sensitivity attainable at a multi-TeV muon collider remains to be explored~\cite{MuonColliderLoI}. 

Finally, we stress the importance of our results for composite Higgs models, where minimal fine-tuning and electroweak precision data point towards a fully-composite right-handed top quark. With the scales of compositeness that we have shown are to be reached, future high-energy colliders will push the concept of naturalness of the electroweak scale to a whole new level, perhaps one where the SM is no longer.\footnote{``You come at the king, you best not miss.'' Omar Devone Little}


\section*{Acknowledgments}

We would like to thank Nick Amin, Aurelio Juste, Jernej Kamenik, Michael Spannowsky, Marcel Vos for helpful discussions, as well as Riccardo Barbieri and Fabio Maltoni for comments on v1. We have been partially supported by the DFG Cluster of Excellence 2094 ORIGINS, the Collaborative Research Center SFB1258 and the BMBF grant 05H18WOCA1, and we thank the MIAPP for hospitality.


\bibliography{futuretops}
\bibliographystyle{jhep}

\end{document}